\documentclass[11pt]{article}
\usepackage[utf8]{inputenc}
\usepackage[T1]{fontenc}
\pdfoutput = 1

\usepackage{amsmath,amssymb,amsthm}
\usepackage{graphicx,subfig}
\usepackage{changepage}
\usepackage{cite,url}
\usepackage{mathtools,xparse,MnSymbol,slashed,mathrsfs}
\usepackage{multirow}
\usepackage{stackrel}
\allowdisplaybreaks
\usepackage{makecell}
\usepackage{stmaryrd}

\usepackage{bbm}
\usepackage{bbold}
\usepackage{caption}
\usepackage{color}
    \definecolor{darkgreen}{rgb}{0,0.5,0}
    \definecolor{darkred}{rgb}{0.5,0,0}
    \definecolor{darkblue}{rgb}{0,0,0.6}
    \definecolor{purple}{rgb}{0.4,.2,0.7}
\usepackage{float}
\usepackage[hyperfootnotes = true, colorlinks = true, linkcolor = darkblue, citecolor = purple]{hyperref}

\numberwithin{equation}{section}

\usepackage[margin = 2.2cm]{geometry}
\usepackage[ragged]{footmisc}
\DeclareMathOperator{\Tr}{Tr}

\newcommand{\clock}{\perp}
\newcommand{\phys}{{\mkern3mu\vphantom{\perp}\vrule depth 0pt\mkern2mu\vrule depth 0pt\mkern3mu}}

\def\bra#1{\langle #1 |}
\def\ket#1{| #1 \rangle}
\def\inner#1#2{\langle #1 | #2 \rangle}

\def\innerc#1#2{\llangle #1 \| #2 \rrangle}

\DeclarePairedDelimiter{\norm}{\lVert}{\rVert}

\usepackage[normalem]{ulem}

\begin{document}

\begin{titlepage}
\thispagestyle{empty}

\begin{flushright}
\end{flushright}

\vspace{1.2cm}
\begin{center}

\noindent{\bf \LARGE Physical Predictions in Closed Quantum Gravity}

\vspace{0.4cm}

{\bf \large Yasunori Nomura$^{a,b,c,d}$ and Tomonori Ugajin$^{e}$}
\vspace{0.3cm}\\

{\it $^a$ Leinweber Institute for Theoretical Physics, Department of Physics, \\
University of California, Berkeley, CA 94720, USA}\\

{\it $^b$ Theoretical Physics Group, Lawrence Berkeley National Laboratory, \\ Berkeley, CA 94720, USA}\\

{\it $^c$ RIKEN Center for Interdisciplinary Theoretical and Mathematical Sciences (iTHEMS), \\
RIKEN, Wako 351-0198, Japan}\\

{\it $^d$ Kavli Institute for the Physics and Mathematics of the Universe (WPI), \\
UTIAS, The University of Tokyo, Kashiwa, Chiba 277-8583, Japan}\\

{\it $^e$ Department of Physics, Rikkyo University, Toshima, Tokyo 171-8501, Japan}

\vspace{0.3cm}
\end{center}

\begin{abstract}
Recent developments in gravitational path integrals indicate that the nonperturbative physical Hilbert space of a closed universe is one-dimensional within each superselection sector. This raises a basic puzzle:\ how can a unique quantum-gravity state give rise to semiclassical physics, measurement outcomes, and classical probabilities? In this paper, we develop a framework in which nontrivial and statistically stable predictions emerge despite the one-dimensionality of the fully constrained Hilbert space. The key idea is to extract physical predictions in an enlarged, unconstrained Hilbert space by conditioning on observational data. We show that partial observability---reflecting the limited access of observers to the degrees of freedom of the universe---suppresses ensemble fluctuations associated with microscopic structure in the gravitational path integral, thereby restoring semiclassical predictability with exponential accuracy. We formulate the construction explicitly including contributions from the Hartle--Hawking no-boundary state, define a gauge-invariant Hilbert space for observations via a density operator, and generalize the formalism to conditioning on histories, clarifying the emergence of classical probabilities and an effective arrow of time. Finally, we explore whether this framework can support a realistic cosmology and identify assumptions that the underlying theory of quantum gravity must satisfy.
\end{abstract}

\end{titlepage}

\tableofcontents
\newpage

\section{Introduction}
\label{sec:intro}

A fundamental question in quantum gravity is how to extract physical predictions in a closed universe, where no external observer or asymptotic boundary is available. In such a setting, diffeomorphism invariance imposes strong constraints on physical states. Recent developments in holographic dualities~\cite{Maldacena:1997re,Ryu:2006bv,Hubeny:2007xt,Faulkner:2013ana,Engelhardt:2014gca} and in gravitational path integrals, including the discovery of replica wormholes~\cite{Penington:2019kki,Almheiri:2019qdq}, have clarified that the nonperturbative physical Hilbert space of a closed universe is one-dimensional within each superselection sector~\cite{Penington:2019npb,Almheiri:2019hni,Usatyuk:2024mzs,Usatyuk:2024isz}. In particular, the Gram matrix constructed from overlaps of candidate states has rank one, implying that all physical states are parallel up to normalization.

At first sight, this result appears to trivialize quantum gravity. If the physical Hilbert space has dimension one, how can the theory describe a rich semiclassical world with nontrivial measurements, decoherence, and probabilistic outcomes? In particular, how can a theory with a unique nonperturbative state account for the apparent multiplicity of classical alternatives observed in cosmology and laboratory experiments? This tension has motivated a number of recent investigations of quantum mechanics and observables in closed universes~\cite{Antonini:2024mci,Harlow:2025pvj,Abdalla:2025gzn,Akers:2025ahe,Chen:2025fwp,Wei:2025guh,Antonini:2025ioh,Harlow:2026hky,Zhao:2026mpl}.

In Ref.~\cite{Nomura:2025whc}, we proposed a resolution of this puzzle based on a different perspective. The key idea is that physical predictions should be extracted not directly in the constrained, nonperturbative Hilbert space, but in an enlarged, {\it unconstrained} Hilbert space in which semiclassical configurations are represented explicitly. Conditioning on observational data in this kinematical Hilbert space produces nontrivial probabilities, even though the underlying physical state is unique. Furthermore, we argued that the statistical fluctuations associated with microscopic ensemble structure in the gravitational path integral are rendered harmless by {\it partial observability}:\ because observers can access only a limited subset of the degrees of freedom in the universe, predictions become statistically stable and semiclassical physics is recovered with exponential accuracy.

The purpose of this paper is to develop this framework in detail and extend it in several directions. First, we formulate the construction explicitly including contributions from the Hartle--Hawking no-boundary state, clarifying the structure of the gravitational inner product and its gauge fixing. Second, we define a gauge-invariant Hilbert space associated with partial observability using the density operator obtained from the gravitational path integral and the Gelfand--Naimark--Segal construction. Third, we generalize the formalism to conditioning on histories rather than single-time data, allowing measurement setups to be implemented consistently within a diffeomorphism-invariant framework and clarifying the emergence of classical probabilities and an effective arrow of time in quantum gravity. Finally, we explore whether this structure can support a realistic cosmology, identifying the assumptions required of the underlying theory of quantum gravity.

Taken together, these results suggest that nonperturbative quantum gravity need not trivialize physical predictions despite the one-dimensionality of the fully constrained Hilbert space. Instead, a nontrivial, stable semiclassical world emerges dynamically once conditioning and partial observability are properly incorporated. The framework developed here provides a coherent setting in which quantum gravity, cosmology, and quantum measurement can be treated within a single unified formalism.

Throughout the paper, we work in a spacetime of dimension $D = d+1$, where $d$ is the number of spatial dimensions.

\section{Quantum Gravity States in an Unconstrained Hilbert Space}
\label{sec:unconstrained}

In quantum gravity, physical states satisfy the Hamiltonian and momentum constraints associated with temporal and spatial diffeomorphism invariance. For semiclassical analyses, however, it is useful to introduce an enlarged, unconstrained Hilbert space ${\cal H}_0$ and to study how physical (gauge-invariant) states are embedded within it.

Let us consider a $d$-dimensional connected compact space ${\cal M}$ and states defined on it.%
\footnote{In this paper, for simplicity we focus on the case where ${\cal M}$ has the topology of a $d$-sphere.}
A convenient basis for the Hilbert space ${\cal H}_{\cal M}$ of these states is labeled by spatial field configurations
\begin{equation}
  q({\bf x}) = \{ h_{ij}({\bf x}), \phi({\bf x}) \},
\end{equation}
consisting of the spatial components of the metric $h_{ij}({\bf x})$ ($i,j=1,\ldots,d$) and matter fields, which we collectively denote by $\phi({\bf x})$. The unconstrained Hilbert space is given by the direct sum of symmetrized products of $n$ ${\cal H}_{\cal M}$ ($n = 0,1,2,\cdots$) corresponding to $n$ disconnected spaces:
\begin{equation}
\begin{aligned}
  {\cal H}_0 &= \bigoplus_{n=0}^\infty \left( {\cal H}_{\cal M}^{\otimes n} \right)_{\rm symm}
\\
  &= {\cal H}_{\emptyset} \oplus {\cal H}_{\cal M} \oplus ({\cal H}_{\cal M} \otimes {\cal H}_{\cal M})_{\rm symm} \oplus \cdots,
\end{aligned}
\end{equation}
where ${\cal H}_{\emptyset} \equiv {\cal H}_{\cal M}^{\otimes 0}$ is the one-dimensional zero-universe sector, spanned by the state $\ket{\emptyset}$, which corresponds in the Lorentzian path integral to a suitable extension of the ``yarmulke'' singularity~\cite{Louko:1995jw}. The Hartle--Hawking ``no-boundary'' state~\cite{Hartle:1983ai} is generated from this state through the gravitational path integral, as described below.

A central object is the ``inner product'' between two states in ${\cal H}_0$. We take this to be a kernel, defined as a group-averaged, time-unoriented Lorentzian gravitational path integral~\cite{Teitelboim:1981ua,Halliwell:1988wc,Halliwell:1990qr} using the canonical variables in the Arnowitt--Deser--Misner (ADM) formalism~\cite{Arnowitt:1959ah}. In this paper, we are mostly interested in such inner products in which one of the two states is a one-boundary state in ${\cal H}_{\cal M}$, because these are the inner products most relevant for making physical predictions at the order of the $e^{-S}$ expansion that we consider. Here, $S$ represents coarse-grained entropies of appropriate systems.

The inner product between no-boundary and one-boundary states is given by the Lorentzian gravitational path integral, with the boundary conditions
\begin{equation}
  h_{ij}({\bf x},t_{\rm o}) = h_{ij}({\bf x}),
\qquad
  \phi({\bf x},t_{\rm o}) = \phi({\bf x})
\label{eq:bc-t_o-1}
\end{equation}
imposed at a {\it coordinate} time $t_{\rm o}$. The path integral includes a yarmulke singularity inserted at another time $t_{\rm i}$, and one must further integrate over $t_{\rm i}$ as well as over the possible yarmulke singularities. The action used for the path integral includes the gauge-fixing term
\begin{equation}
  {\cal I}_{\rm gf} = \int\! d^{d+1}\!x\, \left\{ \Pi\, [\dot{N} - \chi] + \Pi_i [N^i - \chi^i] \right\},
\label{eq:I_gf}
\end{equation}
as well as the corresponding ghost action~\cite{Fradkin:1975cq,Batalin:1983bs}. Here, $N$ and $N^i$ denote the lapse function and shift vector, while $\chi$ and $\chi^i$ are arbitrary functionals of $h_{ij}$, $\phi$, and their conjugate momenta $\pi^{ij}$ and $\pi$. The appearance of $\dot N$ in the gauge-fixing term is required for a consistent variational problem~\cite{Teitelboim:1981ua}. The boundary conditions for variables other than $h_{ij}$ and $\phi$ are given by
\begin{equation}
  \Pi({\bf x},t_{\rm o}) = \Pi_i({\bf x},t_{\rm o}) = c({\bf x},t_{\rm o}) = \bar{c}({\bf x},t_{\rm o}) = 0,
\label{eq:bc-t_o-2}
\end{equation}
with all remaining variables---$\pi^{ij}$, $\pi$, $\rho$, $\bar{\rho}$, $N$, and $N^i$---free at $t_{\rm o}$, except as restricted by the gauge-fixing term; here, $c$ and $\bar{c}$ are the ghosts, with $\rho$ and $\bar{\rho}$ their conjugate momenta. We take the range of integration for the lapse $N$ to include both positive and negative values~\cite{Halliwell:1988wc,Halliwell:1990qr}, the latter corresponding to evolution backward in time (relative to the chosen coordinate time).

We denote the inner product obtained in this way by $G[q({\bf x}); \emptyset]$. Using it, we can construct the following special state in ${\cal H}_{\cal M}$:
\begin{equation}
  \ket{\Psi_{\emptyset}} = \int\! {\cal D}q({\bf x})\, G[q({\bf x}); \emptyset]\,\, \ket{q({\bf x})}.
\label{eq:empty}
\end{equation}
This state---the Hartle--Hawking state restricted to a single connected universe ${\cal M}$---satisfies the Wheeler--DeWitt equation~\cite{Wheeler:1968iap,DeWitt:1967yk}. In other words, it solves the full set of ${\rm Diff}_{d,1}$ constraints.%
\footnote{
 The full Hartle--Hawking no-boundary state represented in ${\cal H}_0$ also has other components, such as $\int\! {\cal D}q_1({\bf x}) {\cal D}q_2({\bf x})\, G[\{ q_1({\bf x}), q_2({\bf x}) \}; \emptyset]\,\, \ket{\{ q_1({\bf x}), q_2({\bf x}) \}}$ in $({\cal H}_{\cal M} \otimes {\cal H}_{\cal M})_{\rm symm}$. We assume that the appropriate symmetry factor is implied in this and analogous expressions.
}

The inner product between states with larger numbers of boundaries can be given similarly. For example, the inner product between two one-boundary states is defined by imposing the boundary conditions in Eqs.~(\ref{eq:bc-t_o-1}) and (\ref{eq:bc-t_o-2}), together with
\begin{equation}
  h_{ij}({\bf x},t_{\rm i}) = \tilde{h}_{ij}({\bf x}),
\qquad
  \phi({\bf x},t_{\rm i}) = \tilde{\phi}({\bf x})
\label{eq:bc_t_i-1}
\end{equation}
and
\begin{equation}
  \Pi({\bf x},t_{\rm i}) = \Pi_i({\bf x},t_{\rm i}) = c({\bf x},t_{\rm i}) = \bar{c}({\bf x},t_{\rm i}) = 0
\label{eq:bc-t_i-2}
\end{equation}
at $t_{\rm i}$, with the remaining variables free, again except as restricted by the gauge-fixing term. The action used for the path integral is the same as before, and the range of integration for $N$ includes both positive and negative values.

The inner product obtained in this way, denoted by $G[q({\bf x}); \tilde{q}({\bf x})]$, is not the {\it kinematical} inner product naturally defined in ${\cal H}_{\cal M}$:
\begin{equation}
  G[q({\bf x}); \tilde{q}({\bf x})] \neq \inner{q({\bf x}) }{ \tilde{q}({\bf x})}.
\end{equation}
Rather, it corresponds, up to an overall normalization, to the group-averaged inner product~\cite{Higuchi:1991tm,Henneaux:1994lbw,Ashtekar:1995zh,Giulini:1998rk,Marolf:2000iq}, formally written as%
\footnote{
 The kinematical inner product between $U(g)\, \ket{q({\bf x})}$ and $U(g)\, \ket{\tilde{q}({\bf x})}$ in ${\cal H}_0$ cannot be used to define a valid inner product either, since $U(g)$ is a distributional operator and hence its square cannot be well defined~\cite{Held:2024rmg}.
}
\begin{equation}
  G[q({\bf x}); \tilde{q}({\bf x})] = \int_{{\rm Diff}_{d,1}}\!\!\! dg\, \bra{q({\bf x})} U(g)\, \ket{\tilde{q}({\bf x})},
\label{eq:GA-inner}
\end{equation}
where $U(g)$ is a (distributional) unitary representation of the {\it spacetime} diffeomorphism group ${\rm Diff}_{d,1}$.%
\footnote{
 The inner product $G[q({\bf x}); \tilde{q}({\bf x})]$ is well defined in ${\cal H}'_{\cal M}$, the Hilbert space obtained by quotienting ${\cal H}_{\cal M}$ by the action of $d$-dimensional spatial diffeomorphisms ${\rm Diff}_d$, in the sense that $G[q_1({\bf x}); \tilde{q}_1({\bf x})] = G[q_2({\bf x}); \tilde{q}_2({\bf x})]$ whenever $q_1({\bf x})$ and $q_2({\bf x})$ are related by a spatial diffeomorphism in ${\rm Diff}_d$, and the same diffeomorphism relates $\tilde{q}_1({\bf x})$ and $\tilde{q}_2({\bf x})$. Here, the spatial diffeomorphism ${\rm Diff}_d$ acts on all spatial slices in a correlated manner, reflecting the gauge redundancy that remains after the gauge fixing in Eq.~(\ref{eq:I_gf}). A similar statement also applies to $G[q({\bf x}); \emptyset]$; i.e., $G[q_1({\bf x}); \emptyset] = G[q_2({\bf x}); \emptyset]$ whenever $q_1({\bf x})$ and $q_2({\bf x})$ are related by ${\rm Diff}_d$.
\label{ft:Diff_d}
}

Using $G[q({\bf x}); \tilde{q}({\bf x})]$ defined this way, we can construct a set of states in ${\cal H}_{\cal M}$, labeled by $\tilde{q}({\bf x})$:
\begin{equation}
  \ket{\Psi_{\tilde{q}({\bf x})}} = \int\! {\cal D}q({\bf x})\, G[q({\bf x}); \tilde{q}({\bf x})]\,\, \ket{q({\bf x})}.
\label{eq:Psi_A}
\end{equation}
These states satisfy the Wheeler--DeWitt equation, meaning that they solve the full set of ${\rm Diff}_{d,1}$ constraints, not just those of ${\rm Diff}_d$. As such, the states $\ket{\Psi_{\tilde{q}({\bf x})}}$, together with $\ket{\Psi_{\emptyset}}$ and analogous states constructed from multi-boundary initial states, form an {\it overcomplete} basis of $\hat{\cal H}_{\cal M}$, the Hilbert space of states on ${\cal M}$ that satisfy both the Hamiltonian and momentum constraints, analogous to continuous, coherent-state-like bases.

The inner product we defined satisfies the standard Hermitian relation:\ for any boundary conditions $A$ and $B$
\begin{equation}
  G[A; B] = (G[B; A])^*.
\label{eq:Hermitian}
\end{equation}
Despite the integration of the lapse $N$ over the whole range including negative values, the ``time reversal'' relation $G[A; B] = G[B; A]$ is not true in general.

In Ref.~\cite{Harlow:2023hjb}, it has been argued that the $CPT$ symmetry must be gauged in quantum gravity. In our language, this corresponds to identifying the boundary conditions $(A,B)$ with $(\bar{B},\bar{A})$ (in the sense of an equivalence class) and projecting states at the level of the kernel by always considering the combination
\begin{equation}
  G[A; B] + G[\bar{B}; \bar{A}] \equiv G_{\rm gauge}[ \llbracket A, B \rrbracket ].
\end{equation}
Here, $\bar{A}$ and $\bar{B}$ represent $CPT$-conjugated states of $A$ and $B$, respectively, and the argument $\llbracket A, B \rrbracket$ of $G_{\rm gauge}$ denotes the equivalence class under $CPT$
\begin{equation}
  \llbracket A, B \rrbracket = \{ (A,B), (\bar{B},\bar{A}) \}.
\label{eq:equiv}
\end{equation}
Because $CPT$ is antiunitary, $CPT$ invariance implies $G[\bar{B}; \bar{A}] = (G[A; B])^*$. We find that $G_{\rm gauge}[ \llbracket A, B \rrbracket ]$ is real and, using Eq.~(\ref{eq:Hermitian}), invariant under interchange of the initial and final boundary conditions, $G_{\rm gauge}[ \llbracket A, B \rrbracket ] = G_{\rm gauge}[ \llbracket B, A \rrbracket ]$. The physical meaning of this gauging is that spacetime histories (or boundary-value problems) related by $CPT$ are identified as physically equivalent. In the rest of the paper, we will specify observational conditions using a representative of the equivalence class in Eq.~(\ref{eq:equiv}); physical predictions are independent of this choice because the gauged theory projects onto $CPT$-invariant boundary data.

\section{Conditioning with Observational Questions}
\label{sec:cond}

In quantum gravity, physical predictions can be made in the form of conditional probabilities. This can be done by projecting the quantum gravitational states onto appropriate states on a conditioning hypersurface representing particular measurement outcomes. In this section, we discuss in detail how conditioning by observational questions is implemented in the current framework.

\subsection{Conditioning in the unconstrained Hilbert space}
\label{subsec:cond-nongrav}

The basic idea is to embed the gravitational state in the {\it unconstrained} (kinematical) Hilbert space ${\cal H}_0$, and to impose in ${\cal H}_0$ the conditions reflecting the measurement setup~\cite{Nomura:2025whc}.

It is useful to remind ourselves that the single-universe Hilbert space ${\cal H}_{\cal M}$ ($\subset {\cal H}_0$) is {\it not} a Hilbert space of ``states at a given time'' in the usual quantum-mechanical sense. This is because, in the absence of a preferred time variable, a configuration $q({\bf x})$ does not by itself correspond to a state at a definite physical time. To assign such an interpretation, one typically chooses one of the fields---for example, a component of $h_{ij}({\bf x})$ or one of the matter fields $\phi({\bf x})$---to serve as a relational time variable, thereby specifying how the hypersurface is embedded in spacetime. Accordingly, we decompose the configuration variables as
\begin{equation}
  q({\bf x}) = \{ q^\clock({\bf x}), q^\phys({\bf x}) \},
\end{equation}
where $q^\clock({\bf x})$ serves as a relational time variable and $q^\phys({\bf x})$ denotes the dynamical fields on equal-time hypersurfaces selected in this way.%
\footnote{
 Subtleties of this decomposition, analogous to the Gribov ambiguity in non-Abelian gauge theories, were discussed in Ref.~\cite{Held:2025mai}. We will not pursue this issue in this paper.
}
We will refer to $q^\clock({\bf x})$, or any degrees of freedom serving the same purpose as a clock, as the clock degrees of freedom.

\begin{figure}[t]
\centering
\vspace{-0.2cm}
  \includegraphics[height=0.3\textwidth]{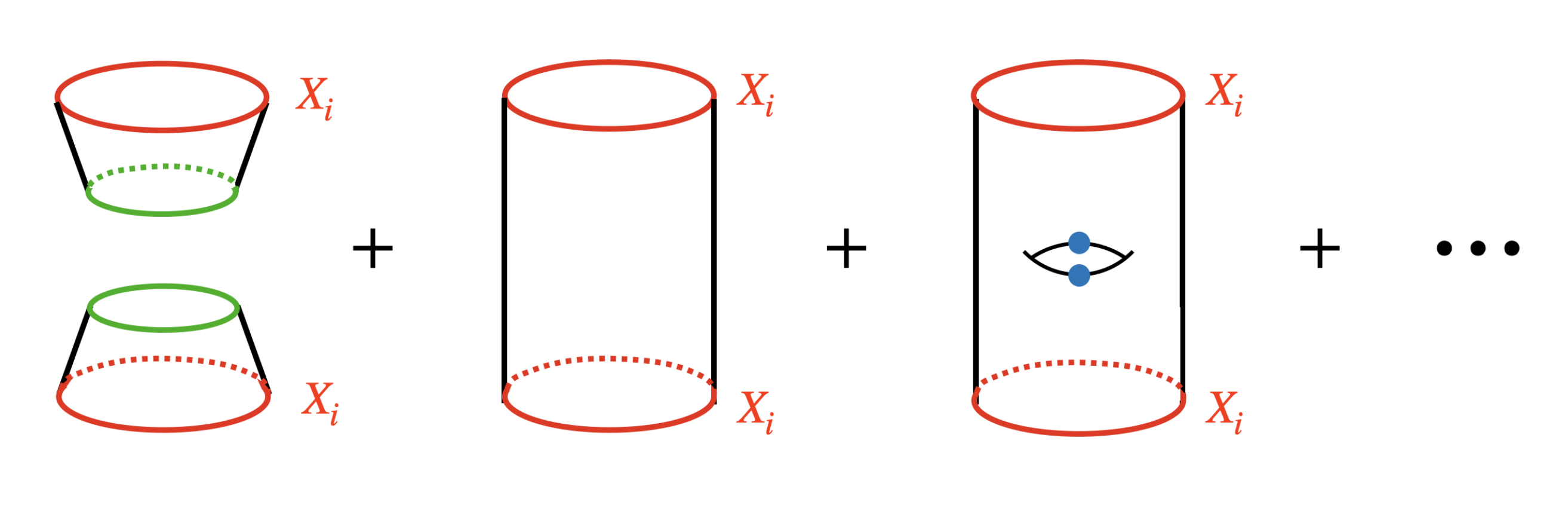}
\vspace{-0.5cm}
\caption{
 Representative gravitational path-integral geometries contributing to the probability of obtaining the measurement outcome $X_i$. Green circles and blue dots represent yarmulke and crotch singularities, respectively.
}
\label{fig:expand}
\end{figure}

Suppose we want to consider the probability $P(X_i)$ of obtaining a particular experimental outcome $X_i$.  For now, we assume that $X_i$ corresponds to {\it a} state in ${\cal H}_0$; this will be generalized in Section~\ref{subsec:semiclassical}. The probability $P(X_i)$ is given by the square of the ``overlap'' ${\cal A}_{\Phi X_i}$ between the quantum gravitational state $\Phi$---nonperturbatively unique~\cite{Penington:2019npb,Almheiri:2019hni,Usatyuk:2024mzs,Usatyuk:2024isz} within each $\alpha$-microsector (discussed further in Section~\ref{subsec:Hilbert})---and the state at some fixed relational time representing $X_i$:%
\footnote{
 This probability is not normalized. It can be used to compute relative probabilities, for example the relative probabilities of obtaining outcomes $X_i$ and $X_j$:\ $P(X_i)/P(X_j)$.
}
\begin{equation}
  P(X_i) = \left| {\cal A}_{\Phi X_i} \right|^2.
\end{equation}
Here, the overlap is understood in the group-averaged sense, and by the overlap with the nonperturbatively unique state $\Phi$, we mean that $P(X_i)$ is calculated by the gravitational path integral without any boundary other than those representing the state $\ket{X_i}$ and its conjugate $\bra{X_i}$. This corresponds to summing the diagrams in Fig.~\ref{fig:expand}.%
\footnote{
 In Ref.~\cite{Nomura:2025whc}, we only considered the diagrams that do not involve yarmulke singularities.
}
As discussed in Section~\ref{sec:partial-obs}, the gravitational path integral implicitly performs an average over the microscopic degrees of freedom corresponding to the $\alpha$-microstates, so $P(X_i)$ should be understood as the corresponding averaged quantity rather than the probability in a fixed $\alpha$-state.

Each diagram in the figure corresponds to a term in the expansion
\begin{equation}
\begin{aligned}
  P(X_i) =& \left| G^{(0)}[X_i;\emptyset] \right|^2 
    + \int\! {\cal D} \tilde{q}({\bf x}) \left| G^{(0)}[X_i; \tilde{q}({\bf x})] \right|^2 
\\
  & + \int\! {\cal D} \tilde{q}_1({\bf x}_1)\, {\cal D} \tilde{q}_2({\bf x}_2) \left| G^{(0)}[X_i; \{ \tilde{q}_1({\bf x}_1), \tilde{q}_2({\bf x}_2) \}] \right|^2 
  + \cdots,
\end{aligned}
\label{eq:P-expand}
\end{equation}
where the superscript $(0)$ means the leading contribution, and $\{ {\bf x} \} = \{ {\bf x}_1 \} \cup \{ {\bf x}_2 \}$ with ${\bf x}_1$ and ${\bf x}_2$ being the coordinates in the two universes appearing in the intermediate stage. 

It is important to note that while Eq.~(\ref{eq:P-expand}) appears to correspond to the insertion of a complete set between $\bra{X_i}$ and $\ket{X_i}$ in the sense of usual quantum mechanics, it in fact does not represent such an insertion. In usual quantum mechanics with a fixed notion of time, a complete set comprises all states at a given time, which would mean that the integrals in Eq.~(\ref{eq:P-expand}) are over the ``physical'' degrees of freedom, $\tilde{q}^\phys({\bf x})$, $\tilde{q}^\phys_1({\bf x}_1)$, $\tilde{q}^\phys_2({\bf x}_2)$, $\cdots$, and not over $\tilde{q}({\bf x})$, $\tilde{q}_1({\bf x}_1)$, $\tilde{q}_2({\bf x}_2)$, $\cdots$. Here we also integrate over the clock data, reflecting the fact that no preferred time is fixed in advance. This reflects that in the theory of gravity, even the size of the Hilbert space for ``physical'' degrees of freedom (Hilbert space for $\tilde{q}^\phys({\bf x})$ with a fixed UV cutoff) changes as ``time'' (e.g., the scale factor of the metric) varies.

Another important point is that probabilities $P(X_i)$---or more precisely, relative probabilities between $X_i$ and $X_j$ for $i \neq j$---are calculated in an unconstrained Hilbert space. $P(X_i)$ computes the square of the coefficients of $\ket{X_i}$---represented as a linear combination of $\ket{q({\bf x})}$ normalized as in usual quantum field theory---when the unique quantum gravitational state is represented in ${\cal H}_0$. We should, therefore, {\it not} normalize $P(X_i)$ by the $X_i$-dependent ``norm'' $G[X_i; X_i]$ computed using the gravitational path integral.%
\footnote{
 While completing this paper, Ref.~\cite{Abdalla:2026mxn} has appeared, in which probabilities are normalized by the gravitational inner product $G[X_i;X_i]$. With this prescription, the states relevant for cosmology become nearly parallel to the Hartle--Hawking state in the Hilbert space defined by the gravitational path integral, so such normalization yields probabilities that are either 1 or nearly 1. In the present framework, by contrast, probabilities are extracted in the unconstrained Hilbert space ${\cal H}_0$, where nontrivial relative probabilities arise once partial observability is taken into account.
}
The path integral already determines the relative weights $P(X_i)$, and hence the relative probabilities $P(X_i)/P(X_j)$. Their common overall normalization is fixed only after specifying a complete set of alternatives, as
\begin{equation}
  P(X_i) \rightarrow \frac{P(X_i)}{\sum_j P(X_j)}.
\end{equation}

\subsection{Relative contributions from varying spacetime topologies}
\label{subsec:relative}

Let us now look into how each diagram in Fig.~\ref{fig:expand} is evaluated more explicitly. Each state (specified, for example, by particle configurations) is generally a superposition of field basis states. So, the first diagram is given by
\begin{equation}
  P^{(\emptyset)}(X_i) = \left| \int\! {\cal D}q^\phys({\bf x})\, f_{X_i}[q^\phys({\bf x})]\, G^{(0)}[q({\bf x}); \emptyset] \right|^2,
\label{eq:P_Xi-0}
\end{equation}
where $f_{X_i}[q^\phys({\bf x})]$ is the weight functional specifying the state $X_i$ on a spacelike hypersurface $\Sigma$ used in imposing observational conditions, which is determined by the clock degrees of freedom $q^\clock({\bf x})$. We assume that weight functionals are normalized such that $\int\! {\cal D}q^\phys({\bf x})\, |f_{X_i}[q^\phys({\bf x})]|^2 = 1$.

In the Lorentzian gravitational path integral, the overlap of a state characterized by $q({\bf x})$ with the no-boundary state is given by the path integral of geometries in which the conditioning boundary (red) in Fig.~\ref{fig:expand} is connected to suitable generalizations of yarmulke singularities (depicted by the green circle in the first diagram):
\begin{equation}
  G^{(0)}[q({\bf x}); \emptyset] = \int\! {\cal D} \tilde{q}({\bf x})\, G[q({\bf x}); \tilde{q}({\bf x})]\, {\cal Y}[\tilde{q}^\phys({\bf x})].
\label{eq:yarmulke-geom}
\end{equation}
Here, ${\cal Y}[\tilde{q}^\phys({\bf x})]$ represents the contribution from a singularity, labeled by $\tilde{q}^\phys({\bf x})$, which is exponentially enhanced by the coarse-grained gravitational entropy of the state specified by $\tilde{q}^\phys({\bf x})$~\cite{Louko:1995jw}
\begin{equation}
  {\cal Y}[\tilde{q}^\phys({\bf x})] \sim e^{S_{\tilde{q}^\phys({\bf x})}}.
\label{eq:yarmulke-enhanc}
\end{equation}
This enhancement alone does not necessarily imply dominance over other diagrams, since none of the relevant geometries may correspond to a stationary point of the action. The suppression coming from rapidly oscillating phases in the path integral would then compete with the effect of this enhancement.

When a stationary point exists, we expect that the corresponding geometry can be extended to a nonsingular, but complex geometry, with the spacetime location of the singularity identified as the hypersurface at which the Lorentzian and Euclidean portions of the geometry are connected (for example, the time at which the radius of the entire universe crosses the de~Sitter radius~\cite{Maldacena:2024uhs}). In this case, the enhancement factor of Eq.~(\ref{eq:yarmulke-enhanc}) can be viewed as arising from the standard Hartle--Hawking no-boundary weighting~\cite{Hartle:1983ai}, associated with the Euclidean portion of the geometry.

The second diagram in Fig.~\ref{fig:expand} is given by
\begin{equation}
  P^{(1)}(X_i) = \int\! {\cal D} \tilde{q}({\bf x})\, \left| \int\! {\cal D}q^\phys({\bf x})\, f_{X_i}[q^\phys({\bf x})]\, G^{(0)}[q({\bf x}); \tilde{q}({\bf x})] \right|^2.
\label{eq:P_Xi-1}
\end{equation}
We may define $G^{(0)}[q({\bf x}); q'({\bf x})]$ on equal clock slices, $q^\clock({\bf x}) = q'^\clock({\bf x})$, by
\begin{equation}
  G^{(0)}[q({\bf x}); q'({\bf x})] = \int\! {\cal D} \tilde{q}({\bf x})\, G^{(0)}[q({\bf x}); \tilde{q}({\bf x})] G^{(0)}[\tilde{q}({\bf x}); q'({\bf x})].
\end{equation}
Using this definition, Eq.~(\ref{eq:P_Xi-1}) can also be written as
\begin{equation}
  P^{(1)}(X_i) = \int\! {\cal D}q^\phys({\bf x})\, f_{X_i}[q^\phys({\bf x})] \int\! {\cal D}q'^\phys({\bf x})\, f_{X_i}^*[q'^\phys({\bf x})]\, G^{(0)}[q({\bf x}); q'({\bf x})] \Bigr|_{q^\clock({\bf x}) = q'^\clock({\bf x})}.
\end{equation}

The third diagram in Fig.~\ref{fig:expand} can be written as
\begin{equation}
  P^{(2)}(X_i) = \int\! {\cal D} \tilde{q}_1({\bf x}_1)\, {\cal D} \tilde{q}_2({\bf x}_2) \left| \int\! {\cal D}q^\phys({\bf x})\, f_{X_i}[q^\phys({\bf x})]\, G^{(0)}[q({\bf x}); \{ \tilde{q}_1({\bf x}_1), \tilde{q}_2({\bf x}_2) \}] \right|^2.
\end{equation}
In the Lorentzian signature, the quantity $G^{(0)}[q({\bf x}); \{ \tilde{q}_1({\bf x}_1), \tilde{q}_2({\bf x}_2) \}]$, represented by a trouser-shaped geometry, generically involves a ``crotch''-type singularity~\cite{Louko:1995jw} (depicted by blue dots in the figure), where a single connected spatial region splits into two disconnected components. Unlike the yarmulke case, the contribution of this singularity is exponentially suppressed rather than enhanced:
\begin{equation}
  G^{(0)}[q({\bf x}); \{ \tilde{q}_1({\bf x}_1), \tilde{q}_2({\bf x}_2) \}] \sim e^{-S_{\rm crotch}}.
\end{equation}
Here, $S_{\rm crotch}$ should be understood as an effective action cost associated with topology change, rather than as a thermodynamic entropy. We expect $S_{\rm crotch}$ to be parametrically large when two universes resulting from the split are larger than the Planck scale.

We note that if the cutoff scale $M_*$ of the low-energy effective theory is taken to be the Planck scale $M_{\rm Pl}$ (which is not possible if the number of low-energy species $N_{\rm eff}$ is parametrically larger than $O(1)$~\cite{Dvali:2007hz}), off-shell contributions from geometries of different topologies would vary continuously across topology changes in the path integral. For example, contributions to $P^{(\emptyset)}(X_i)$ from geometries with ``small'' singularities, $S_{\tilde{q}^\phys({\bf x})} \sim O(1)$, would be continuously connected to those contributing to $P^{(1)}(X_i)$, in which the cylinder is tightly cinched at some height. The relative path-integral suppression of adding a ``tiny'' handle, converting, e.g., a geometry in $P^{(1)}(X_i)$ to that in $P^{(2)}(X_i)$, would also be of order $e^{-S_{\rm crotch}}$, which is $O(1)$ when $S_{\rm crotch} \sim O(1)$. This continuity, however, appears to be lost if there is a separation between the cutoff and Planck scales, by effects of order $e^{-O(M_{\rm Pl}^{d-1}/M_*^{d-1})} \sim e^{-O(N_{\rm eff})}$.

\begin{figure}[t]
\centering
\vspace{0cm}
  \includegraphics[height=0.27\textwidth]{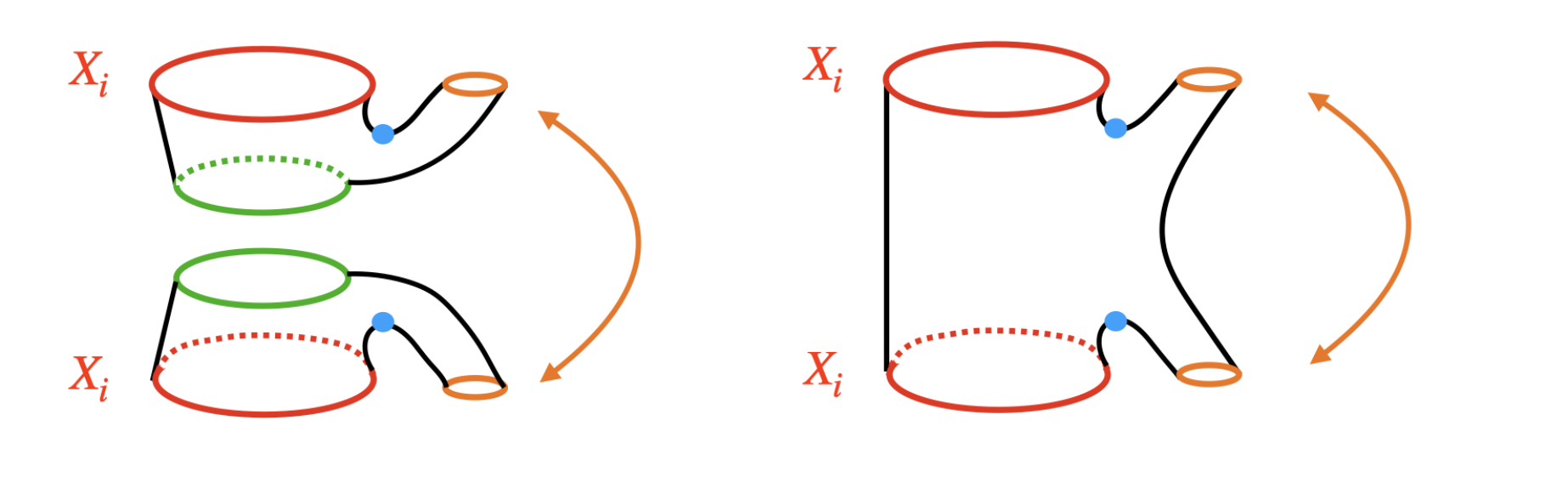}
\vspace{-0.3cm}
\caption{
 Examples of conditioning geometries containing disconnected universes. Disconnected components on the conditioning hypersurface that carry no observational data are traced out.}
\label{fig:disconn}
\end{figure}

Finally, we note that the conditioning hypersurface may contain disconnected universes in which no observational condition is imposed. Since these universes are not observed, they must be traced out (see later). Examples of this tracing out are depicted in Fig.~\ref{fig:disconn}.
These diagrams, however, are exponentially suppressed compared with the corresponding diagrams without disconnected universes, by powers of factors of the form $e^{-S_{\rm crotch}}$. We will not discuss the contributions from these diagrams further in this paper.%
\footnote{
 Diagrams which are disconnected from the observation region {\it in spacetime} do not contribute to relative probabilities as vacuum bubbles do not contribute to correlation functions in usual quantum field theory.
}

\subsection{Gauge fixing in the bulk and conditioning hypersurface}
\label{subsec:gauge-fix}

It is useful to analyze how gauge fixing is implemented in Eq.~(\ref{eq:yarmulke-geom}) for $P^{(\emptyset)}(X_i)$ and Eq.~(\ref{eq:P_Xi-1}) for $P^{(1)}(X_i)$. Once the ``initial'' hypersurface, $t = t_{\rm i}$, is specified by $\tilde{q}^\clock({\bf x})$ in the chosen coordinate system, subsequent constant coordinate-time hypersurfaces are determined in a way consistent with the bulk gauge fixing in Eq.~(\ref{eq:I_gf}). Suppose we take a simple gauge
\begin{equation}
  \chi = \chi^i = 0.
\end{equation}
In this case, the spatial coordinates in subsequent constant coordinate-time hypersurfaces are all determined in terms of the coordinates on the initial hypersurface by the condition $N^i = 0$. The condition $\dot{N} = 0$ fixes time reparameterization, except for the residual position-dependent time rescaling of the form $t \rightarrow \zeta({\bf x})\, t$; the position-dependent time shift is fixed by the introduction of the initial hypersurface at $t = t_{\rm i}$.

Observational conditions, such as $X_i$, are imposed on a fixed spacelike hypersurface $\Sigma$. The physical hypersurface $\Sigma$ is determined by specifying the configuration $q^\clock({\bf x})$. On the other hand, this hypersurface must lie on the constant coordinate-time slice, $t = t_{\rm f}$. This partially fixes the gauge associated with the embedding of $\Sigma$. In particular, the gauge transformation parameters $\xi^\mu$ near $\Sigma$ must take the form
\begin{equation}
  \xi^0(t_{\rm f},{\bf x}) = 0,
\qquad
  \partial_t \xi^0(t_{\rm f},{\bf x}): \text{fixed by } q^\clock({\bf x}),
\end{equation}
\begin{equation}
  \xi^i(t_{\rm f},{\bf x}): \text{arbitrary},
\qquad
  \partial_t \xi^i(t_{\rm f},{\bf x}): \text{fixed to keep } N^i = 0,
\end{equation}
where $\partial_t \xi^0(t_{\rm f},{\bf x})$ is fixed by the above condition. Note that this fixing of $\partial_t \xi^0$ on $\Sigma$ eliminates entirely the gauge redundancies associated with the position-dependent time rescaling, $t \rightarrow \zeta({\bf x})\, t$, remaining after imposing $\dot{N} = 0$. The fixing of $\partial_t \xi^i(t_{\rm f},{\bf x})$ comes from the condition $N^i = 0$, together with its preservation under time evolution. After all these gauge-fixing restrictions, the only residual gauge redundancies are those associated with $\xi^{i}(t_{\rm f},{\bf x})$, which are the spatial diffeomorphisms ${\rm Diff}_d$ discussed, e.g., in footnote~\ref{ft:Diff_d}. This structure is preserved after we integrate over $\tilde{q}^\clock({\bf x})$.

The gauge fixing for a spacetime geometry with splitting/merging universes, such as that for $P^{(2)}(X_i)$, can be analyzed analogously. A subtlety is that, under a splitting/merging, the correspondence between the coordinates ${\bf x}$ of the universe before the splitting (after the merging) and those ${\bf x}_1$ and ${\bf x}_2$ of the two universes after the splitting (before the merging) is not strictly one-to-one. This subtlety, however, is restricted to a measure-zero subset of the coordinates, and we do not worry about it in this paper.

In any case, we find that the introduction of $\Sigma$ reduces ${\rm Diff}_{d,1}$ to a smaller subset. The conditioning used to make a physical prediction must be phrased in terms of the quantities ``on $\Sigma$'':\ $q^\clock({\bf x})$, which specifies $\Sigma$, and $f_{X_i}[q^\phys({\bf x})]$, which provides the measurement condition by assigning amplitudes to field configurations and (through a functional Fourier transform) their conjugate momenta. Physical predictions are therefore obtained by phrasing the observational condition entirely in terms of these quantities. Because the bulk fields are fully integrated over in the gravitational path integral defining the gauge-invariant kernel, these predictions do not depend on the choice of bulk gauge fixing, as discussed in Section~\ref{sec:unconstrained}.

\subsection{Conditioning and spacetime diffeomorphisms}
\label{subsec:cond-viol}

Since the introduction of the hypersurface $\Sigma$ amounts to a partial gauge fixing, the remaining spacetime diffeomorphism invariance is realized somewhat nontrivially. In particular, the quantities $q^\clock({\bf x})$ and $f_{X_i}[q^\phys({\bf x})]$ specifying the conditioning are not independent. Changing $q^\clock({\bf x})$ changes the physical hypersurface $\Sigma$, so the conditioning functional $f_{X_i}[q^\phys({\bf x})]$ must be modified accordingly in order to represent the same physical observational condition. Physical predictions, however, depend only on this combined relational specification of the conditioning and, as discussed in Section~\ref{subsec:gauge-fix}, are independent of the choice of bulk gauge fixing.

This structure of correlation between $q^\clock({\bf x})$ and $f_{X_i}[q^\phys({\bf x})]$ is reminiscent of formulations in which explicit ``observer'' degrees of freedom are added to the theory~\cite{Harlow:2025pvj,Abdalla:2025gzn,Akers:2025ahe,Chandrasekaran:2022cip,Balasubramanian:2023xyd,Geng:2020qvw,Geng:2024dbl}, where the gravitational constraints are imposed on the combined system of observers and spacetime. In our framework, specifying the clock degrees of freedom plays the role of erecting such an observer and conditioning on the configurations of $q^\phys({\bf x})$ corresponds to describing the physical system relative to that observer:\ the theory of quantum gravity, when properly interpreted, already contains all the ingredients needed to make meaningful predictions without introducing additional external degrees of freedom.

Another important aspect of the present framework is the crucial difference between gravity and ordinary gauge theories. In a non-gravitational gauge theory, such as a confining Yang--Mills theory, observers external to a process can be treated as singlets under the gauge symmetry. One may therefore formulate physical questions entirely in terms of gauge-invariant operators, without reference to gauge-dependent quantities. In gravity, by contrast, spacetime coordinates themselves are gauge, and any operational description of observations necessarily refers to configurations expressed in a particular coordinate system. In this sense, the way we specify the questions we pose necessarily involves a choice of coordinate representation, even though the resulting physical predictions are gauge invariant. Accordingly, the conditioning that defines these predictions cannot be formulated within the fully constrained (gauge-invariant) Hilbert space; it necessarily requires the enlarged, unconstrained Hilbert space in which the coordinate-dependent data specifying the question are represented explicitly.

\section{Ensembles, {\boldmath $\alpha$}-States, and Partial Observability}
\label{sec:partial-obs}

The gravitational theory considered here is nonrenormalizable and hence should be viewed as a low-energy effective theory below some cutoff scale $M_*$ (the string scale, if the UV theory is string theory). Nevertheless, it has been shown that the gravitational path integral including spacetime wormholes captures certain aspects of the UV theory, such as the unitarity of black hole evolution as viewed from the exterior~\cite{Penington:2019kki,Almheiri:2019qdq} and the exact number of microstates of Bogomol'nyi--Prasad--Sommerfield (BPS) black holes~\cite{Iliesiu:2022kny}.

The statistical nature emerging from such path integrals is closely related~\cite{Nomura:2025whc} to the existence of $\alpha$-states~\cite{Coleman:1988cy,Giddings:1988cx,Marolf:2020xie}, which arise from small, UV wormholes. While this statistical nature potentially jeopardizes semiclassical physics~\cite{Harlow:2025pvj,Abdalla:2025gzn}, robust semiclassical predictions are recovered (up to exponentially small corrections) through {\it partial observability}~\cite{Nomura:2025whc}---the fact that we cannot observe the entire universe. In this section, we study how these aspects appear in the framework developed so far. We will see that partial observability can dramatically change the dominance structure of the gravitational path integral.

The $\alpha$-state description used here should not be understood as requiring $\alpha$ parameters to be fundamental ingredients of the exact UV theory. Rather, it provides a description of the ensemble structure that can emerge in the semiclassical gravitational path integral. If the exact theory does not contain such a fundamental ensemble structure, the conditioning framework developed in this paper is unaffected. Our discussion below shows that even when such an ensemble structure appears in the semiclassical description, its effects on physical predictions are exponentially suppressed.

\subsection{Microscopic ensembles from {\boldmath $\alpha$}-states}
\label{subsec:alpha}

As we have seen, the amplitude of emitting or absorbing a baby universe (a separate universe) is suppressed by a factor of $e^{-S_{\rm crotch}}$. Since the action cost scales with the size of the baby universe, this factor is largest when the baby universe is smallest, i.e.\ of order the cutoff size.

In the low-energy effective theory, the creation and annihilation of such small baby universes can be described by contributions to the Hamiltonian density of the form~\cite{Coleman:1988cy,Giddings:1988cx}
\begin{equation}
  {\cal H}_{\rm baby} = \sum_i {\cal O}_i[ q({\bf x}) ,p({\bf x}) ]\, A_i,
\end{equation}
where $p({\bf x}) = \{ \pi^{ij}({\bf x}), \pi({\bf x}) \}$ represents conjugate momenta for $q({\bf x})$, and ${\cal O}_i[ q({\bf x}) ,p({\bf x}) ]$ are taken, without loss of generality, to be a linearly independent set of Hermitian functionals. The operators $A_i$ are position-independent ``baby universe operators,'' satisfying
\begin{equation}
  A_i = A_i^\dagger,
\qquad
  [A_i, A_j] = 0.
\label{eq:baby-ops}
\end{equation}

Given the properties in Eq.~(\ref{eq:baby-ops}), states in the low-energy effective theory are decomposed into the so-called $\alpha$-states---a set of states defining superselection sectors for observables within a single universe. States in different $\alpha$-sectors correspond to eigenstates of $A_i$ with different eigenvalues, which can be treated as continuous parameters in the low-energy semiclassical description. This does not imply that the underlying microscopic states in a complete theory of quantum gravity form a continuum; they may instead be discrete, with $\alpha$ providing a coarse-grained semiclassical label. This amounts to considering an ensemble of effective field theories with varying field content, masses, and coupling constants, which we collectively denote by $\alpha$. The effect of small, UV wormholes can thus be represented by a probability density function ${\cal P}(\alpha)$ defined at the cutoff scale.

Because the parameters $\alpha$ form a continuous set, a physical measurement performed during a finite time interval cannot determine their precise values. Predicting the outcome of such a measurement therefore involves averaging over underlying microscopic degrees of freedom that cannot be resolved at the semiclassical level and that correspond, at the classical level, to $\alpha$; we refer to these as $\alpha$-microstates. This averaging is implicitly incorporated in the gravitational path integral over semiclassical degrees of freedom, arising as a coarse-grained effect of integrating out the microscopic sector.

\subsection{Emergence of semiclassical physics}
\label{subsec:semiclassical}

\subsubsection*{Appearance of the ensemble nature in the low-energy effective theory}

The ensemble nature resulting from $\alpha$-microstates is visible in the low-energy effective theory through a class of spacetime wormholes called replica wormholes~\cite{Penington:2019kki,Almheiri:2019qdq}.

\begin{figure}[t]
\centering
\vspace{0.1cm}
  \includegraphics[height=0.38\textwidth]{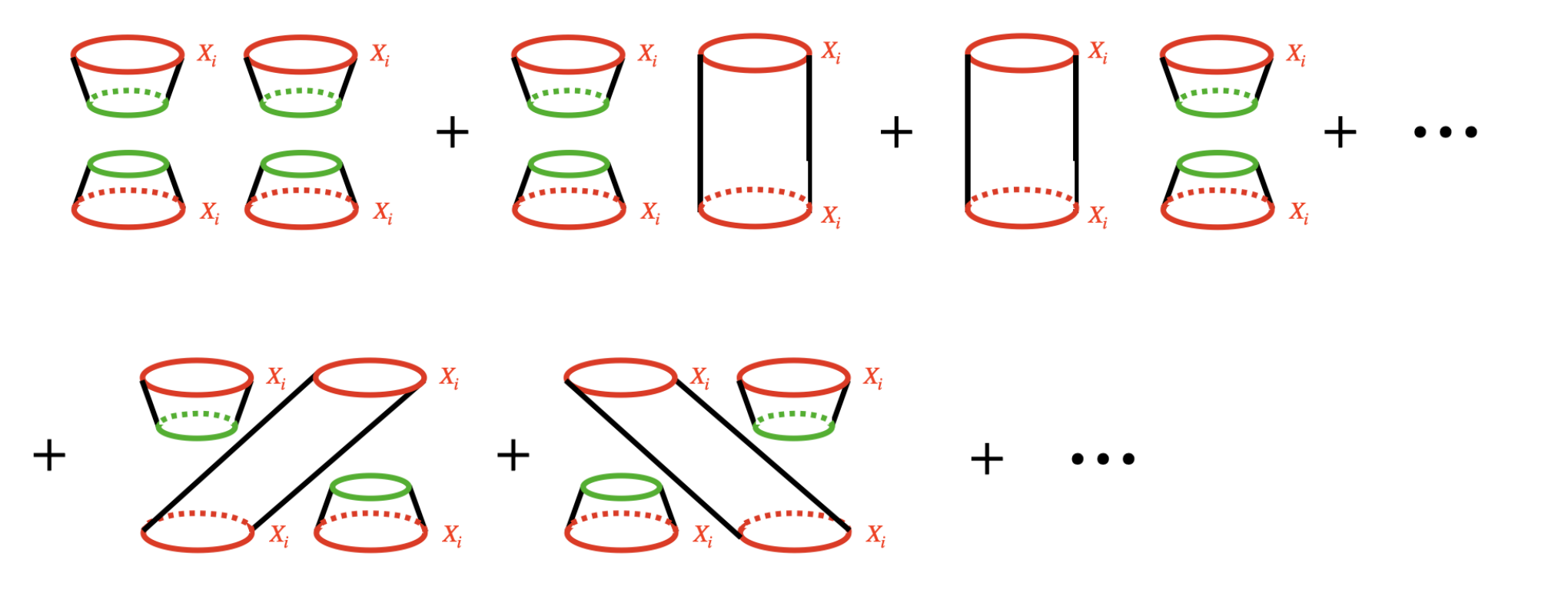}
\vspace{-0.9cm}
\caption{
 Saddles contributing to the gravitational path integral for $P(X_i)^2$. The first line shows contributions corresponding to the square of the saddles for $P(X_i)$, while the second line gives contributions that do not have corresponding contributions in the computation of $P(X_i)$.
}
\label{fig:nb-square}
\end{figure}

Suppose there is a saddle in which the universe satisfying condition $X_i$ is born from the no-boundary state. In this case, the probability $P(X_i)$ can be calculated by the gravitational path integral depicted in Fig.~\ref{fig:expand}. On the other hand, to calculate $P(X_i)^2$, we need to perform a gravitational path integral with a doubled number of boundaries. This provides extra contributions beyond the square of the contribution in Fig.~\ref{fig:expand}, as depicted in the second line of Fig.~\ref{fig:nb-square}.
This discrepancy is attributed to the fact that the gravitational path integral computes ensemble averages---in this case over $\alpha$-microstates---so that $\overline{P(X_i)^2}$ can be different from $\bigl( \overline{P(X_i)} \bigr)^2$~\cite{Saad:2019lba,Stanford:2019vob}, where the overline represents the ensemble average. In the case considered here, the fractional variation over the ensemble is of order
\begin{equation}
  \frac{\sqrt{\overline{P(X_i)^2} - \bigl( \overline{P(X_i)} \bigr)^2}}{\overline{P(X_i)}} \sim e^{-S_{\rm yarmulke}},
\label{eq:yarm-supp}
\end{equation}
where $S_{\rm yarmulke}$ denotes the (coarse-grained) entropy associated with the yarmulke singularity (the de~Sitter entropy for a de~Sitter state emerging through the Hartle--Hawking no-boundary solution).

\begin{figure}[t]
\centering
\vspace{-0.3cm}
  \includegraphics[width=\textwidth]{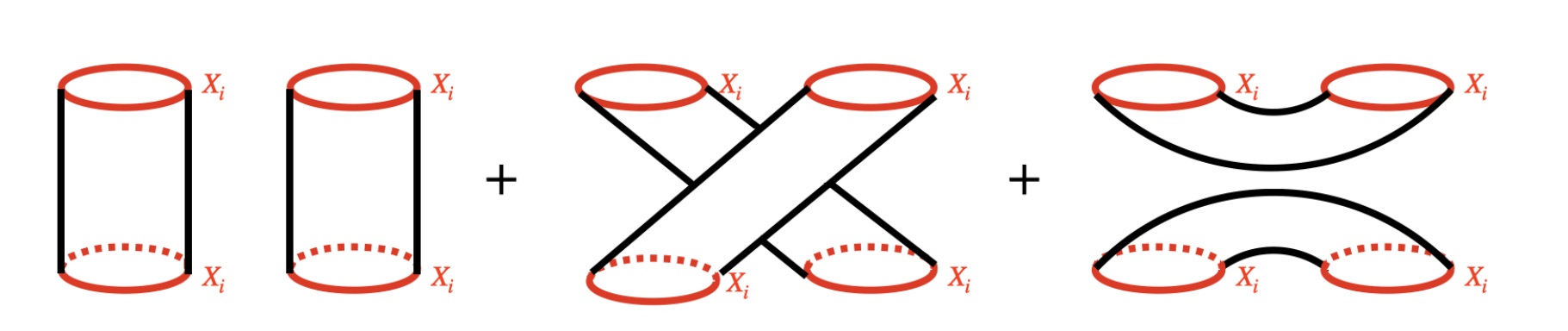}
\vspace{-0.7cm}
\caption{
 Saddles contributing to the gravitational path integral for $P(X_i)^2$. All three have equal magnitude, but only the first equals the square of the saddle for $P(X_i)$.
}
\label{fig:cyl-square}
\end{figure}

A more dramatic situation occurs if there is no saddle involving the no-boundary state and the probability is dominated by the ``cylinder'' geometry, i.e.\ the second diagram in Fig.~\ref{fig:expand}. In this case, the gravitational path integral computing $P(X_i)^2$ receives three leading contributions shown in Fig.~\ref{fig:cyl-square}.
While only the first diagram corresponds to the square of the saddle computing $P(X_i)$, all three have equal magnitude, leading to~\cite{Harlow:2025pvj}
\begin{equation}
  \frac{\sqrt{\overline{P(X_i)^2} - \bigl( \overline{P(X_i)} \bigr)^2}}{\overline{P(X_i)}} \simeq \sqrt{2}.
\label{eq:error}
\end{equation}
This implies that the value of $P(X_i)$ fluctuates strongly---fractionally of order unity---from one $\alpha$-microstate to another. Consequently, in this case, predictions made by conditioning $X_i$ on the state of the entire universe are not meaningful unless the $\alpha$-microstate is known exactly, which is impossible in a finite interval of time.

\subsubsection*{Recovery of semiclassicality from partial observability}

There is, however, a good reason why this issue does not pose a problem for physical predictions. While large fluctuations of predictions across $\alpha$-microstates would be problematic, these fluctuations are suppressed by partial observability:\ the fact that the microscopic state of the entire universe cannot be operationally determined, even in principle~\cite{Nomura:2025whc}.

Suppose the system under consideration is located in a spatial region $R$; i.e., the clock degrees of freedom are specified only in region $R$, and the weight functional $f_{X_i}[q^\phys({\bf x})]$ has support only in $R$ representing the configurations of the system there. In this case, fields outside $R$ are path integrated without imposing any observational conditions, so that
\begin{equation}
\begin{aligned}
  P(X_i) \sim& \int\! {\cal D} \tilde{q}({\bf x})\, 
  \left\{ \int_{{\bf x} \in R} {\cal D}q^\phys({\bf x})\, {\cal D}q^{\prime\phys}({\bf x})\, f_{X_i}[q^\phys({\bf x})]^* f_{X_i}[q^{\prime\phys}({\bf x})] \right\}
\\
  & \quad\times 
  \left\{ \int_{{\bf x} \notin R} {\cal D}q({\bf x})\, {\cal D}q'({\bf x})\, \delta[q({\bf x}) - q'({\bf x})] \right\} 
  G[\tilde{q}({\bf x}); q({\bf x})]\, G[q'({\bf x}); \tilde{q}({\bf x})].
\end{aligned}
\label{eq:P_Xi}
\end{equation}
Note that the ``gluing'' outside $R$ is purely a kinematical tracing in ${\cal H}_0$ and does not involve an additional gravitational path integral; consequently it does not generate gravitational wormholes. Also, outside $R$, all degrees of freedom---including clock variables---are traced over. This expresses $P(X_i)$ as a Schwinger--Keldysh-type gluing of forward and backward gravitational histories along the same conditioning hypersurface outside $R$~\cite{Hartle:2010dq,Hartle:2016tpo,Ivo:2024ill}.

We emphasize that this tracing is an ordinary partial trace in the kinematical Hilbert space ${\cal H}_0$. For a fixed observational condition $X_i$ supported in $R$, one could equivalently retain the degrees of freedom outside $R$ and insert the identity operator on them, obtaining the same result. The distinction from the discussion leading to Eq.~(\ref{eq:error}) is that there $X_i$ specifies a fine-grained condition on the entire universe, whereas here configurations outside $R$ are summed over, defining an inclusive, coarse-grained observational condition. While the partial trace itself does not alter the probability for this fixed condition, the ensemble fluctuations of the resulting inclusive probability can differ from those of the fine-grained probabilities. If additional degrees of freedom outside $R$ are observed, the conditioning is simply refined to include them, with only the remaining unobserved degrees of freedom traced out.

In fact, to satisfy observational condition $X_i$, we need not fully specify the microstate in $R$. By denoting weight functionals satisfying the condition $X_i$ by $f^a_{X_i}[q^\phys({\bf x})]$ ($a = 1,2,\cdots$), the expression in Eq.~(\ref{eq:P_Xi}) is modified to%
\footnote{
 For a fixed outcome $X_i$, the functions $f^a_{X_i}$ form an orthonormal set:\ $\int\! {\cal D}q^\phys({\bf x})\, f^{a*}_{X_i}[q^\phys({\bf x})]\, f^{a'}_{X_i}[q^\phys({\bf x})] = \delta_{a a'}$.
}
\begin{equation}
\begin{aligned}
  P(X_i) \sim& \int\! {\cal D} \tilde{q}({\bf x})\, 
  \left\{ \sum_{a,a'} \int_{{\bf x} \in R} {\cal D}q^\phys({\bf x})\, {\cal D}q^{\prime\phys}({\bf x})\, f^a_{X_i}[q^\phys({\bf x})]^* f^{a'}_{X_i}[q^{\prime\phys}({\bf x})] \right\}
\\
  & \quad\times 
  \left\{ \int_{{\bf x} \notin R} {\cal D}q({\bf x})\, {\cal D}q'({\bf x})\, \delta[q({\bf x}) - q'({\bf x})] \right\} 
  G[\tilde{q}({\bf x}); q({\bf x})]\, G[q'({\bf x}); \tilde{q}({\bf x})].
\end{aligned}
\label{eq:PXi-tr}
\end{equation}
Here, spatial continuity of the fields is implicit (unless there is an appropriate distributional source). We denote this quantity by
\begin{equation}
  \int\! {\cal D} \tilde{q}({\bf x})\, G[\tilde{q}({\bf x}); (X_i,X_i); \tilde{q}({\bf x})] \equiv \rho_{ii},
\label{eq:rho_Xi}
\end{equation}
which can be viewed as a Schwinger--Keldysh-type gluing of all the non-conditioned degrees of freedom under $X_i$. As can be checked using Eq.~(\ref{eq:Hermitian}), it is real:\ $\rho_{ii}^* = \rho_{ii}$.

\begin{figure}[t]
\centering
\vspace{0.1cm}
  \includegraphics[width=0.97\textwidth]{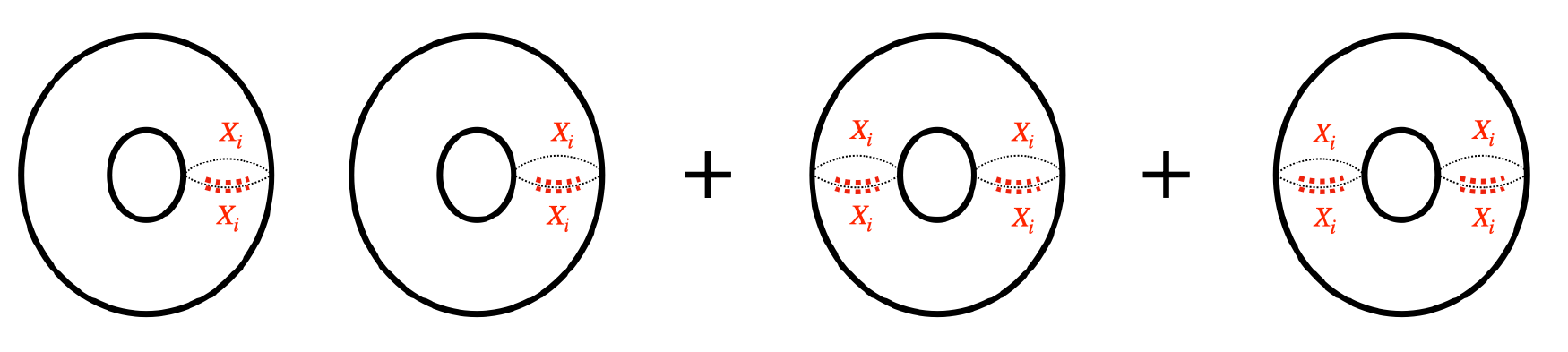}
\vspace{-0.2cm}
\caption{
 Saddles contributing to the gravitational path integral for $P(X_i)^2$ when conditioning is restricted to a subset of the whole degrees of freedom. The topologies of the first and other diagrams differ. As a result, $P(X_i)^2$ is dominated by the first saddle, suppressing replica-wormhole contributions.
}
\label{fig:cyl-square-tr}
\end{figure}

The coarse graining implicit in this observational condition changes the topology of the leading-order cylindrical diagrams contributing to the ensemble moments from those in Fig.~\ref{fig:cyl-square} to those in Fig.~\ref{fig:cyl-square-tr}.
This makes the first diagram, which represents the square of $\overline{P(X_i)}$, dominate over others~\cite{Nomura:2025whc}. The first diagram receives two contributions (in each connected component) from ``loops'' of traced-out matter degrees of freedom, which enhance the total contribution of the diagram by a factor of $(e^{S_{\rm unobs}})^2$. On the other hand, the second and third diagrams each receive only one such contribution:\ $e^{S_{\rm unobs}}$. Here, $S_{\rm unobs}$ corresponds to the logarithm of the number of independent states in $\hat{\cal H}_{\cal M}$---the Hilbert space of states in the effective theory which satisfy the ${\rm Diff}_{d,1}$ constraints---and not the coarse-grained thermodynamic entropy of the traced-out degrees of freedom evaluated on the conditioning hypersurface. This suppresses the variance of the prediction over $\alpha$-microstates:
\begin{equation}
  \frac{\sqrt{\overline{P(X_i)^2} - \bigl( \overline{P(X_i)} \bigr)^2}}{\overline{P(X_i)}} \sim e^{-\frac{1}{2}S_{\rm unobs}},
\label{eq:partial-suppr}
\end{equation}
restoring semiclassicality.

If there is a saddle involving yarmulke singularity, the deviation from semiclassicality is already suppressed, as in Eq.~(\ref{eq:yarm-supp}). The contribution of matter provides a factor of order $e^{O(1) S_{\rm unobs}}$ for each connected diagram in Fig.~\ref{fig:nb-square}.%
\footnote{
 As an example of a component involving a yarmulke singularity, one can consider the case in which a slow-roll inflating universe emerges from the Hartle--Hawking no-boundary state. The factor $O(1) S_{\rm unobs}$ in this case corresponds (for a particular choice of $R$) to $\ln\rho_{\rm fa} - \ln\rho_c = (3/2-2\ln 2) \epsilon_r S_r$ in Ref.~\cite{Ivo:2024ill}, where $\epsilon_r$ and $S_r$ are the slow-roll parameter and the de~Sitter entropy, respectively. This is, indeed, smaller than $S_{\rm yarmulke} = S_r$ by a factor $\epsilon_r \ll 1$.
}
Since the number of degrees of freedom relevant for the matter contribution is generally smaller than that for the gravitational contribution, $S_{\rm unobs} < S_{\rm yarmulke}$, this contribution does not change the fact that semiclassical physics is obtained with exponential accuracy.

\section{Hilbert Space, Physical Predictions, and Conditioning on Histories}
\label{sec:predictions}

In this section, we develop a framework for extracting physical predictions from quantum gravity when full observability of the universe is neither possible nor required. While the exact nonperturbative Hilbert space of quantum gravity is highly constrained, meaningful physical predictions arise once we condition on a restricted set of outcomes accessible to observers. This naturally leads to an effective, gauge-invariant Hilbert space associated with partial observability, together with a well-defined notion of expectation values and probabilities.

We first explain how a nontrivial low-energy Hilbert space emerges from conditioning on incomplete information, using the gravitational path integral to construct a density operator and the associated Hilbert space. We then show how classical probabilities and measurement outcomes arise dynamically through decoherence, clarifying how particular experiments are selected within a theory describing the entire universe. Finally, we generalize the discussion to conditioning on histories rather than single-time data, formulating physical predictions in terms of decoherent histories in quantum gravity and elucidating the emergence of an effective arrow of time. We conclude by explaining how the resulting framework reproduces familiar flat-space quantum mechanics and $S$-matrix physics in the appropriate limits.

\subsection{Gauge-invariant low-energy Hilbert Space from partial observability}
\label{subsec:Hilbert}

It is by now well-established that the nonperturbative physical Hilbert space of quantum gravity is one-dimensional within each $\alpha$-microsector. This can be seen by computing the Gram matrix, defined as overlaps $M_{a a'} = {\cal A}_{\Psi_a \Psi_{a'}}$ among a complete set of states $\Psi_a$ in ${\cal H}_0$, using the gravitational path integral including replica wormhole contributions. The computation gives
\begin{align}
  & \overline{\Tr(M^n)} = \overline{(\Tr M)^n},
\\
  & \overline{(\Tr(M^n) - (\Tr M)^n)^2} = 0,
\end{align}
where overlines denote averaging over $\alpha$-microstates. This implies that the Gram matrix satisfies
\begin{equation}
  \Tr(M^n) = (\Tr M)^n
\end{equation}
for each member of the ensemble. Since the Gram matrix is positive semidefinite and has nonzero trace, the only matrices satisfying this relation are rank-one matrices. Therefore, the physical Hilbert space in each $\alpha$-microsector is one-dimensional.

A nontrivial Hilbert space can, however, be generated through partial observability. To this end, we derive the density operator $\rho$ {\it defined on} the vector space spanned by the conditioning data $X_i$ ($i = 1,2,\cdots$), where each $X_i$ denotes a possible outcome used for conditioning and need not correspond to a fully decohered measurement outcome. With partial observability, the deviation from semiclassicality is exponentially suppressed, as discussed in Section~\ref{subsec:semiclassical}, so we neglect it henceforth.

\begin{figure}[t]
\centering
\vspace{0.3cm}
  \includegraphics[height=0.25\textwidth]{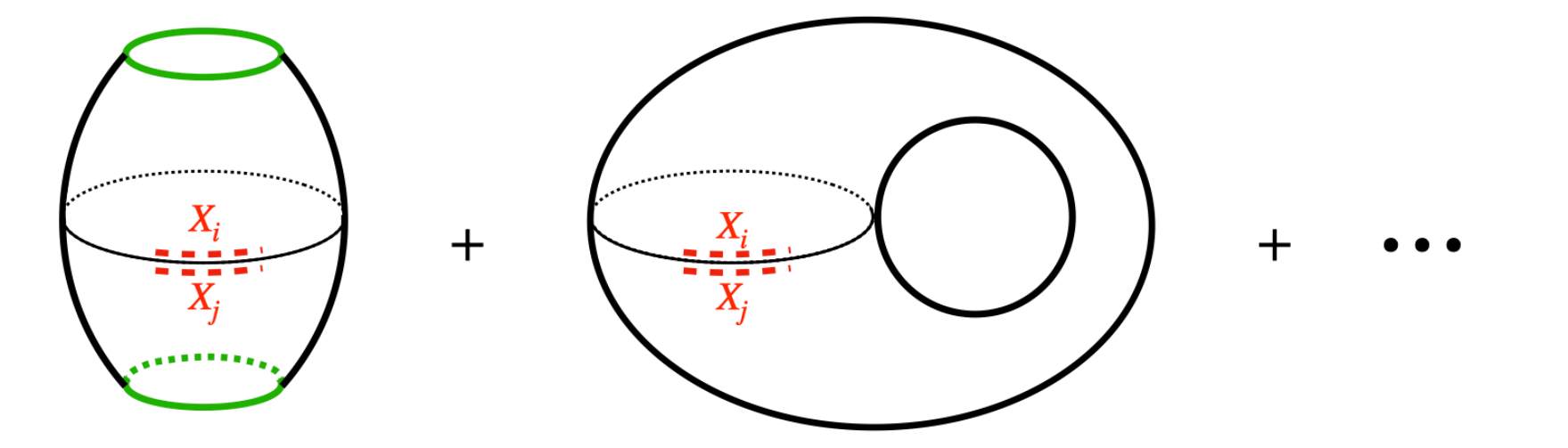}
\vspace{0cm}
\caption{
 Gravitational path integral giving the density operator $\rho_{ij}$. Representative saddles contributing to $\rho_{ij}$, obtained by replacing one of the boundary conditions $X_i$ in the computation of $P(X_i)$ with $X_j$, are shown. Contributions include no-boundary and cylindrical geometries, with degrees of freedom not constrained by $X_i$ or $X_j$ traced out.
}
\label{fig:density-op}
\end{figure}

The elements of $\rho$ can be calculated using the gravitational path integral, where one of the $X_i$ boundary conditions appearing in the computation of $P(X_i)$ is replaced by $X_j$~\cite{Page:1986vw,Hawking:1986vj}; see Fig.~\ref{fig:density-op}.
For example, when a saddle involving the no-boundary state exists, we obtain
\begin{equation}
\begin{aligned}
  \rho_{ij} \sim& \sum_{a,a'} \int_{{\bf x} \in R} {\cal D}q^\phys({\bf x})\, {\cal D}q^{\prime\phys}({\bf x})\, f^a_{X_i}[q^\phys({\bf x})]^* f^{a'}_{X_j}[q^{\prime\phys}({\bf x})]
\\
  & \quad\times 
  \left\{ \int_{{\bf x} \notin R} {\cal D}q({\bf x})\, {\cal D}q'({\bf x})\, \delta[q({\bf x}) - q'({\bf x})] \right\} 
  G[\emptyset; q({\bf x})]\, G[q'({\bf x}); \emptyset] + \cdots.
\end{aligned}
\end{equation}
In the absence of such a saddle, the leading contribution comes from the second diagram in Fig.~\ref{fig:density-op}, giving Eq.~(\ref{eq:PXi-tr}) with $f^{a'}_{X_i}[q^{\prime\phys}({\bf x})]$ replaced with $f^{a'}_{X_j}[q^{\prime\phys}({\bf x})]$. Note that the ranges of the sums over $a,a',\cdots$ may differ for different outcomes $X_i$. As before, $\rho$ is not normalized; physical density matrices are obtained by dividing by ${\rm Tr}\rho$.

Operators in the $X_i$ space can be expressed as
\begin{equation}
  \Psi = \sum_{a,a'} \int_{{\bf x} \in R} {\cal D}q^\phys({\bf x})\, {\cal D}q^{\prime\phys}({\bf x})\, f^a_{X_i}[q^\phys({\bf x})]\, \ket{q({\bf x})}\, \Psi_{ij}\, \bra{q'({\bf x})}\, f^{a'}_{X_j}[q^{\prime\phys}({\bf x})]^*,
\label{eq:GNS-op}
\end{equation}
forming the operator algebra acting on the outcome space. We restrict attention to operators supported in the observed region $R$, so operator differences localized entirely outside $R$ are absent from the algebra.

Given the density operator and the algebra, we can adopt the Gelfand--Naimark--Segal (GNS) construction~\cite{Gelfand:1943,Segal:1953,Haag:1996hvx} to erect the Hilbert space associated with it. In this construction, operators are first identified if their difference is invisible to the state $\rho$, i.e.\ if
\begin{equation}
  {\rm Tr}\!\left[\rho\,(\Psi-\Phi)^\dagger(\Psi-\Phi)\right] = 0.
\end{equation}
Vectors in the Hilbert space are equivalence classes of operators under this identification. On these equivalence classes, an inner product is then defined by
\begin{equation}
  \innerc{\Psi}{\Phi} = \frac{{\rm Tr}[\rho\, \Psi^\dagger \Phi]}{{\rm Tr}\rho} = \frac{\sum_{i,j,k} \rho_{ij} \Psi^\dagger_{jk} \Phi_{ki}}{\sum_i \rho_{ii}}.
\end{equation}
The density operator $\rho$ provides a distinguished cyclic state (corresponding to the equivalence class of the identity operator), and expectation values of observables ${\mathcal O} = {\mathcal O}^\dagger$, i.e.\ ${\mathcal O}$ given as in Eq.~(\ref{eq:GNS-op}) with ${\mathcal O}_{ij} = {\mathcal O}_{ji}^*$, are given by
\begin{equation}
  \langle \mathcal O \rangle = \frac{{\rm Tr}[\rho\, \mathcal O]}{{\rm Tr}\rho}.
\end{equation}

While the Hilbert space obtained in this way depends on the choice of $\{ X_i \}$, it is gauge invariant in the appropriate sense. As discussed in Section~\ref{subsec:cond-viol}, adopting a different gauge in the bulk requires corresponding changes in $q^\clock({\bf x})$ and $f^a_{X_i}[q^\phys({\bf x})]$ so that they are associated with the same outcome $X_i$. Being defined on the conditioning hypersurface, operators $\Psi$ must also be adjusted accordingly. With these transformations (of $q^\clock({\bf x})$, $f^a_{X_i}[q^\phys({\bf x})]$, and $\Psi$), the resulting Hilbert spaces in different gauges are physically equivalent, in the sense that they yield identical expectation values for all observables associated with the same outcomes.

It is important to stress that this construction is not obtained by simply selecting a subspace of the kinematical Hilbert space ${\cal H}_0$. A naive restriction of ${\cal H}_0$ would in general depend on the choice of coordinates and would therefore fail to define a gauge-invariant Hilbert space. The gauge invariance of the present construction relies crucially on the correlated transformation of $q^\clock({\bf x})$, $f^a_{X_i}[q^\phys({\bf x})]$, and operators $\Psi$ under changes of bulk gauge, as discussed in Section~\ref{subsec:cond-viol}.

\subsection{Emergence of classical probabilities from a measurement}
\label{subsec:classical}

\subsubsection*{Selecting a particular experiment}

When we make the choice of $\{ X_i \}$, it is important to note that the state we are considering is that for {\it the entire universe}. For example, if $X_i$ is chosen just to specify an electron arriving at a particular location on a screen of a double-slit experiment, the conditioning selects {\it all experiments} performed in the universe that are consistent with the conditioning. The resulting probabilities then receive contributions from all the experiments, some of which may include a which-slit detector and some of which may not.

To discuss the consequence of a particular experiment, we thus have to include in the outcomes $X_i$ information that is sufficient to identify the specific experiment (or possibly a collection of experiments consistent with our knowledge). For example, we need to include in $\{ X_i \}$ the fact that we perform our experiment at a particular location on Earth, in the form of a superposition of field configurations. This condition about the experimental setup can be---and indeed is typically---common to all possible outcomes $X_i$, and plays the role of selecting the measurement context. The density operator obtained in this way then corresponds to outcomes of the particular experiment.

\begin{figure}[t]
\centering
\vspace{-0.3cm}
  \includegraphics[height=0.4\textwidth]{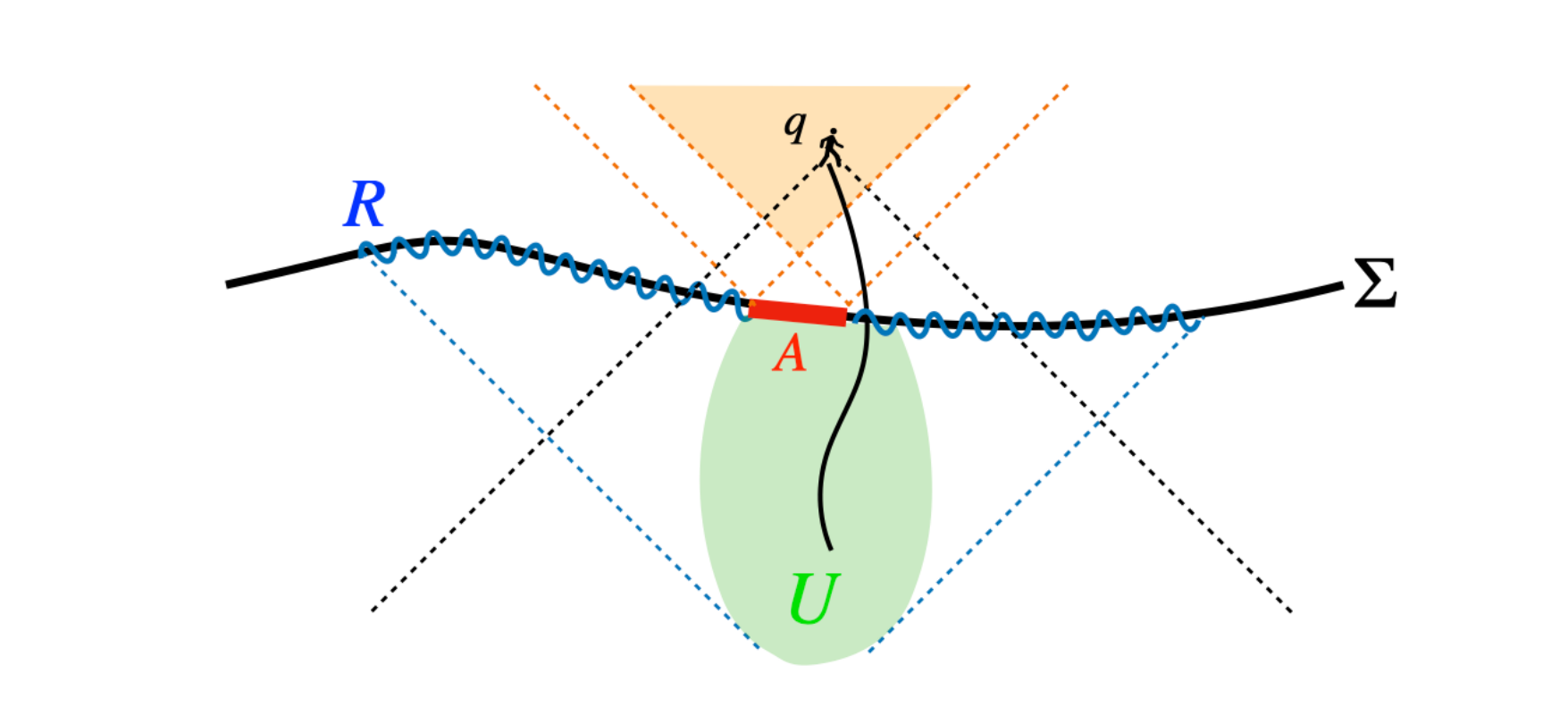}
\vspace{-0.2cm}
\caption{
 Spacetime regions relevant for conditioning on a particular experiment. The experimental apparatus after the measurement is localized in the region $A$ on the conditioning hypersurface $\Sigma$. An observer who can access the outcome must lie in the spacetime region $O = \cap_{p \in A} J^+(p)$ (yellow). The observer's available records are restricted to a region $U \subset J^-(q)$ (green), and the conditioning associated with the experimental setup is imposed on $R = J^+(U) \cap \Sigma$.
}
\label{fig:exp-setup}
\end{figure}

In order to see how this procedure can be implemented, suppose that the experimental apparatus after the measurement is located in the region $A$ on a conditioning spacelike hypersurface $\Sigma$ (specified by the configuration of the clock degrees of freedom); see Fig.~\ref{fig:exp-setup}.
Let us assume, for simplicity, that the physical observer performing the experiment needs to collect information from all parts of the apparatus. An observer with knowledge of the outcome must be located in the spacetime region
\begin{equation}
  O = \bigcap_{p \in A} J^+(p)
\end{equation}
(the yellow region in the figure). Here, $J^+(p)$ refers to the causal future of $p$.

Now, the observer, located at a spacetime point $q \in O$, can, in principle, have access to information from the causal past of $q$, $J^-(q)$. In the present formulation, however, conditioning data are defined on a spacelike hypersurface $\Sigma$, so only information whose records are present on $\Sigma$ can be included in the conditioning. Moreover, even within $J^-(q) \cap J^-(\Sigma)$, an observer can in practice access only a limited spacetime region $U$, typically surrounding the past trajectory of the observer (the green region in the figure). The conditioning associated with the observer's knowledge---i.e.\ the experimental setup---is therefore imposed on the region
\begin{equation}
  R = J^+(U) \cap \Sigma
\end{equation}
of the conditioning hypersurface. With this definition, events that occur in the future of $A$ but whose records are created only after $\Sigma$ are not included in the conditioning, even if they lie in the causal past of the observer at $q$. In general, it is not easy to express the knowledge about $U$ purely on $R$, though in a semiclassical saddle this can be done using equations of motion~\cite{Nomura:2011dt}. For now, we assume that this can be done and study the consequences, deferring discussion of a more direct way to impose this conditioning to the next subsection.

\subsubsection*{Emergence of classical realities}

The tracing-out procedure determines the structure of the resulting density operator on the outcome space. Suppose we have $n$ outcomes $X_i$ ($i = 1,2,\cdots,n$). The density operator may then be represented as an $n \times n$ matrix in the outcome space (an unnormalized version of the density matrix). We can normalize this matrix by dividing it by its trace, which gives the density matrix $\tilde{\rho}_{ij}$ for the outcomes.

The rank of the density matrix $\tilde{\rho}_{ij}$ indicates the degree to which the outcomes $X_i$ are entangled with the traced-out degrees of freedom on the conditioning hypersurface $\Sigma$, and hence the extent to which the measurement has decohered~\cite{Zurek:1981xq,Joos:1984uk} with its environment. As is well known, this entanglement generically develops on short timescales, and it increases the rank of the density matrix.

If the rank of the density matrix is full, ${\rm rank}(\tilde{\rho}) = n$, and the matrix is approximately diagonal in a stable basis, the measurement can be viewed as fully decohered. We can then select a special basis $Y_a$ ($a = 1,2,\cdots,n$) in which
\begin{equation}
  \tilde{\rho}_{ab} \approx P(Y_a)\, \delta_{ab}.
\end{equation}
The outcomes $Y_a$ can then be identified as classical realities arising from the measurement, selected dynamically by environment-induced superselection (``einselection'')~\cite{Zurek:2003zz}. In the fully decohered regime, the diagonal elements $P(Y_a)$ admit an interpretation as classical probabilities corresponding to the emergent classical realities.

If the interactions between the degrees of freedom representing the outcomes and the remaining degrees of freedom are weak in the past of $\Sigma$, the rank of $\tilde{\rho}_{ij}$ may be much smaller than $n$. In particular, if the rank is approximately unity, in the sense that one eigenvalue $P(Y_1)$ is much larger than all others, then the system is well approximated by a pure state. In this regime, the outcome space has not yet split into distinct classical alternatives, and the dynamics is effectively described by a single quantum branch evolving unitarily in the GNS Hilbert space constructed in Section~\ref{subsec:Hilbert}.

\subsection{Conditioning on histories in quantum gravity}
\label{subsec:history}

So far, we have assumed that all observational conditions, including those specifying the measurement setup, are imposed on an ``equal-time'' hypersurface $\Sigma$. It is, however, useful to allow the conditioning data---such as the initial configuration of the measurement---to be imposed at earlier times within the causal past of the observer.

For this purpose, we adopt the framework of consistent histories~\cite{Griffiths:1984rx,Omnes:1988ek,Gell-Mann:1992wkv,Hartle:1992}, in which quantum theory is formulated as a theory of probabilistic histories, rather than of instantaneous measurement outcomes. In this framework, a history is defined by a sequence of sets of events, say $\alpha \equiv (\alpha_1, \alpha_2, \cdots, \alpha_n)$, occurring at a series of times $t_1 < t_2 < \cdots < t_n$. In a theory with a fixed notion of time, this can be represented by the corresponding chain of appropriately coarse-grained Heisenberg-picture projection operators
\begin{equation}
  C_\alpha = {\cal P}^{(n)}_{\alpha_n}(t_n) \,\cdots\, {\cal P}^{(2)}_{\alpha_2}(t_2)\, {\cal P}^{(1)}_{\alpha_1}(t_1).
\end{equation}
Given an initial state $\ket{\Psi}$, the branch state corresponding to the history $\alpha$ is defined by $C_\alpha \ket{\Psi}$, and the decoherence functional between two histories $\alpha$ and $\beta$ is defined as%
\footnote{
 More generally, for an initial density matrix $\tilde{\rho}$, the decoherence functional is $D(\alpha,\beta) = \mathrm{Tr}[C_\beta \, \tilde{\rho} \, C_\alpha^\dagger]$.
}
\begin{equation}
  D(\alpha,\beta) = \bra{\Psi} C_\alpha^\dagger C_\beta \ket{\Psi}.
\end{equation}
This complex number measures the degree of quantum interference between the two histories:\ if $D(\alpha,\beta) \neq 0$ for $\alpha\neq\beta$, the histories interfere and cannot be assigned classical probabilities. 
A set of histories $\{ \alpha, \beta, \cdots \}$ is said to be consistent if
\begin{equation}
  D(\alpha,\beta) \approx 0 \quad \text{for all } \alpha\neq\beta,
\end{equation}
in which case classical probabilities satisfying the Kolmogorov sum rules, $p(\alpha \vee \beta) = p(\alpha) + p(\beta)$, can be assigned:%
\footnote{
 Strictly speaking, it is only necessary that $\mathrm{Re}[D(\alpha,\beta)] \approx 0$ for $\alpha \neq \beta$ in order for consistent classical probabilities to be assigned. In realistic cases of decoherent histories, however, $D(\alpha,\beta) \approx 0$ for $\alpha\neq\beta$, and we assume this to be the case for simplicity.
}
\begin{equation}
  P(\alpha) = D(\alpha,\alpha) = \norm{C_\alpha \ket{\Psi}}^2.
\label{eq:D-clas}
\end{equation}

\begin{figure}[t]
\centering
\vspace{0.3cm}
  \includegraphics[height=0.39\textwidth]{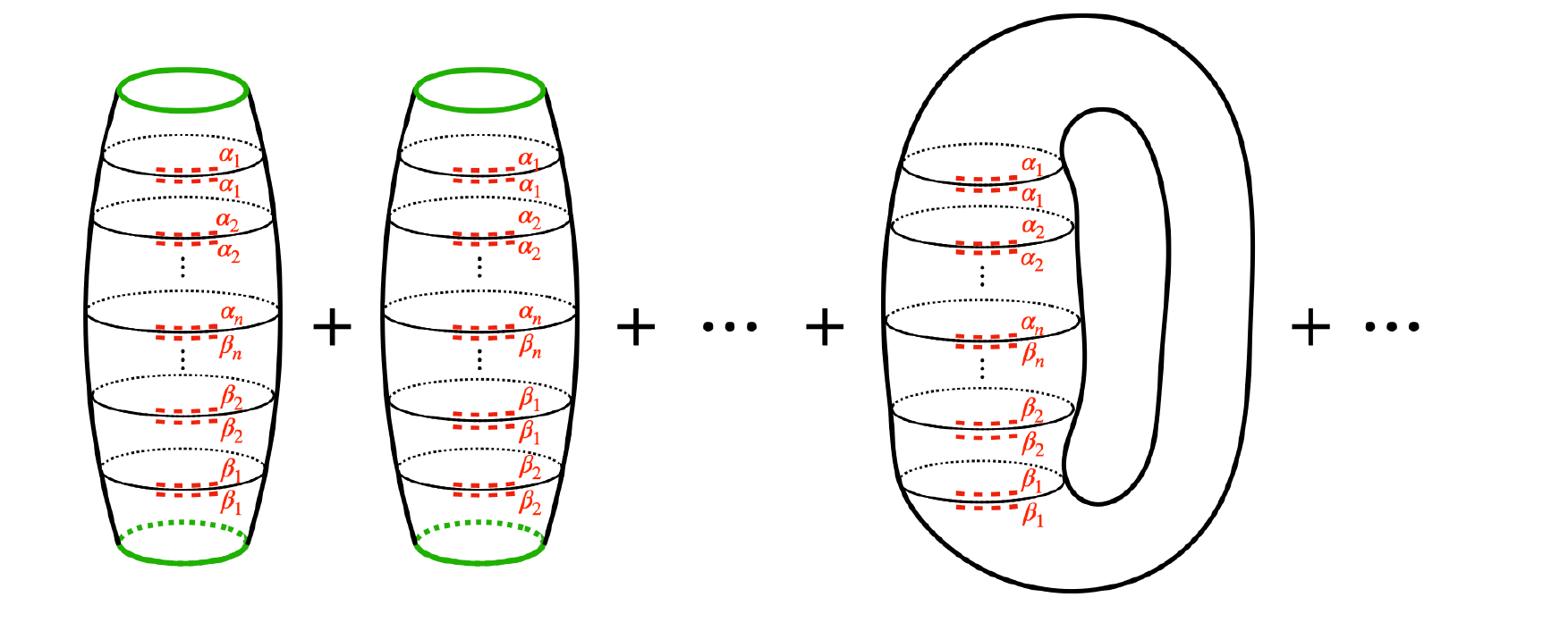}
\vspace{-0.2cm}
\caption{
 Representative saddle-point contributions to the decoherence functional $D(\alpha,\beta)$, with boundary conditions specified by observational data and relational clock variables. Shown are:\ the leading-order saddle involving the no-boundary state with boundary conditions ordered consistently with the thermodynamic arrow of time; representative saddles with incorrect orderings of boundary conditions (shown explicitly and by ellipses), which are exponentially suppressed; and a next-to-leading-order contribution not involving the no-boundary state, together with additional incorrectly ordered and higher-order saddles (indicated by ellipses). Only saddles with the correct ordering admit semiclassical solutions and dominate the decoherence functional.
}
\label{fig:dec-func}
\end{figure}

In a theory of quantum gravity, a state of a closed system does not depend on time in the usual sense; physical time evolution emerges only relationally, through correlations among physical degrees of freedom~\cite{DeWitt:1967yk,Page:1983uc}. As we have been doing, physical time can be specified by configurations of clock degrees of freedom. This implies that the decoherence functional can be calculated as a gravitational path integral, with observational conditions on various hypersurfaces specified by relational clock data, as depicted in Fig.~\ref{fig:dec-func}.
Using the notation in Eq.~(\ref{eq:rho_Xi}), the leading-order contribution to the decoherence functional is given, when a saddle involving the no-boundary state exists, by
\begin{equation}
\begin{aligned}
  D(\alpha,\beta) &\sim G[\emptyset; (\alpha_1,\alpha_1); (\alpha_2,\alpha_2); \cdots; (\alpha_{n-1},\alpha_{n-1}); (\alpha_n,\beta_n); (\beta_{n-1},\beta_{n-1}); \cdots; (\beta_2,\beta_2); (\beta_1,\beta_1); \emptyset]
\\
  &\quad + \mbox{permutations},
\end{aligned}
\label{eq:decoh-grav}
\end{equation}
up to overall normalization, where the ordered sequence of paired boundary conditions implements the insertion of $C_\alpha^\dagger C_\beta$ in the gravitational path integral. Here, we have assumed, for simplicity, that alternative histories are specified using the same number of time steps, $n$. We can further assume, naturally, that the series of times are common between $\alpha$ and $\beta$; i.e., the portions of the hypersurfaces on which $(\alpha_i, \alpha_i)$ and $(\beta_i, \beta_i)$ boundary conditions are imposed are specified by the same clock data (for each $i = 1,\cdots,n-1$). This alignment ensures that overall normalization and gauge-volume factors cancel automatically in the final expressions.

The permutations in Eq.~(\ref{eq:decoh-grav}) consist of terms obtained by permuting the boundary data $(\alpha_i, \alpha_i)$, $(\beta_i, \beta_i)$, and $(\alpha_n, \beta_n)$, together with the associated clock data. Since there is no intrinsic notion of time---or of a preferred time flow---in quantum gravity, these terms are not a priori excluded and therefore contribute to the decoherence functional. However, if the ordering of boundary data does not correspond to the usual thermodynamic arrow of time (e.g., when the coarse-grained entropy fails to increase monotonically in sufficiently isolated regions), the contribution is suppressed.

This can be seen at the level of the saddle-point approximation as follows. By choosing the gauge $\dot{N} = N^i =0$, for example, the sign of $N$ is fixed along the worldline of a point with fixed spatial coordinates. Configurations with ``wrong orderings'' therefore cannot satisfy the equations of motion, since the lapse cannot reverse sign along a classical trajectory---barring the highly unlikely possibility that the boundary conditions are genuinely realized in such orders, as might occur in exotic cyclic cosmologies. As a result, such contributions are exponentially suppressed in the saddle-point approximation.

With the decoherence functional obtained in this way, relative probabilities of obtaining various classical worlds can be calculated through Eq.~(\ref{eq:D-clas}). Here, histories $\alpha, \beta, \cdots$ correspond to outcomes $X_i$ ($i = 1,2,\cdots$) in the previous prescription of conditioning everything on a single time slice $\Sigma$. The treatment here, however, makes the emergence of the concept of the flow of time more manifest. If the histories are taken to be compatible with measurement-like situations, the probabilities $P(\alpha)$ reproduce the Born rule, with exponential accuracy.%
\footnote{
 There are two distinct limiting ways to implement conditioning on the measurement setup. One is to impose conditions successively in small spatial regions on hypersurfaces near points along the worldline (or geodesic) of the physical observer. The other is to deform the conditioning hypersurface $\Sigma$ toward the past so that it approaches the boundary of the causal past of the observer, imposing all conditions there at once. These two approaches are reminiscent of the ``fat-geodesic''~\cite{Bousso:2008hz} and ``causal patch''~\cite{Bousso:2006ev} measures, respectively, in the sense of the spacetime regions considered. Here, however, these constructions are used solely to specify the observational setup, not to define a probabilistic measure by counting events within those regions. The probabilities arise strictly from quantum mechanics through conditioning in the Hilbert space, as envisioned in Ref.~\cite{Nomura:2011dt}.
}

\subsubsection*{Relation to flat space physics}

If the curvature scale of spacetime is much larger than the characteristic length scale of a process (or equivalently, if the curvature is small compared to the characteristic energy scale), the process can be well approximated as occurring in flat space. Furthermore, if the universe is much larger than the system, deviations from semiclassical physics are expected to be exponentially suppressed by a factor of $e^{-S_{\rm univ}}$, where $S_{\rm univ}$ is the smallest of $S_{\rm yarmulke}$, $S_{\rm unobs}/2$, and similar contributions; see Eqs.~(\ref{eq:yarm-supp}) and (\ref{eq:partial-suppr}). In the limit of an infinitely large universe, $S_{\rm univ}$ is expected to diverge, implying that the effects of $\alpha$-microstates become physically irrelevant, consistent with the baby-universe hypothesis of Refs.~\cite{Marolf:2020xie,McNamara:2020uza}.%
\footnote{
 This implies that, in this limit, there is no need to introduce superselected sectors beyond the reach of observables. If one further assumes the existence of physical processes connecting all possible vacua, this suggests that the cobordism group of quantum gravity formulated in an asymptotically flat background is trivial.
}

As discussed in Section~\ref{subsec:classical}, if the system associated with the process is sufficiently isolated from the rest of the degrees of freedom, its evolution can be described (approximately) by unitary time evolution in the Hilbert space describing it. Suppose we choose two physical times to describe histories, one far in the past and the other far in the future relative to the region where the process is occurring (by taking the configurations of clock degrees of freedom as such). Then, the amplitude
\begin{equation}
  U_{ij} \sim G\bigl[\emptyset; (X^F_i,X^F_i); (X^I_j,X^I_j); \emptyset\bigr] + \cdots
\end{equation}
describes the transition amplitude which, after appropriate normalization, defines a unitary $S$-matrix. Here, we have assumed that indices $i,j,\cdots$ exhaust all configurations relevant for the process.

Realistically, this approximate $S$-matrix paradigm is what underlies the experiments we perform in practice. The standard formalism based on asymptotic in- and out-states provides an idealized description of this regime.

\section{Toward a Realistic Cosmology}
\label{sec:cosmo}

The framework described in this paper provides a setting to explore a cosmology that does not require an additional ingredient such as a ``theory of initial conditions.'' While the framework is based on an effective field theory, and hence cannot predict certain quantities such as probability distributions of parameters, we can consider a model that adopts assumptions about their values or distributions.%
\footnote{
 At a fundamental level, probability distributions for these parameters may be determined by the string landscape~\cite{Bousso:2000xa,Susskind:2003kw,Douglas:2003um}.
}
In this final section, we discuss whether it is possible to construct a successful cosmology along these lines, and, if so, what one would have to assume about the properties of the underlying fundamental theory of quantum gravity.

A virtue of the present framework is that probabilities for outcomes of a measurement are determined by quantum mechanics without referring to the global spacetime structure (as anticipated in earlier work~\cite{Nomura:2011dt,Nomura:2011rb,Garriga:2012bc,Friedrich:2022tqk}). This avoids various paradoxes (such as the ``youngness'' paradox) that arise when probabilities are weighted by global spacetime volume~\cite{Guth:2000ka,Tegmark:2004qd,Feldstein:2005bm,Garriga:2005ee,Page:2009qe}. There are, however, still important issues that must be addressed in order to obtain a realistic cosmology.

\subsubsection*{Problem of too large curvature}

We assume that our universe went through a phase of slow-roll inflation driven by the energy density of one or more scalar fields (inflatons) at an early stage. In this framework, it is known that if the universe is dominantly created via the no-boundary mechanism (represented in our context by a yarmulke singularity), then it predicts too large a spatial curvature of the universe~\cite{Maldacena:2024uhs,Vilenkin:1987kf}. Unfortunately, this preference for larger spatial curvature is so strong that it seems unlikely to be countered by environmental selection alone, i.e.\ just by anthropic arguments.

In order for sufficient inflation to occur, the universe must be created at some point in the potential sufficiently higher than the endpoint of slow-roll inflation (not necessarily along the slow-roll trajectory of the inflaton). The amplitude for creating a universe in such a way is expected to be smaller than that for directly producing a slow-roll inflating universe, since the coarse-grained (de~Sitter) entropy (the value of $S_{\rm yarmulke}$) associated with such a higher point is smaller than that associated with the endpoint of slow-roll inflation. To avoid this issue within the present framework, we would thus have to assume the following:
\begin{itemize}
\item[(1)]
The universe cannot be created directly in its slow-roll phase.
\end{itemize}

An interesting possibility to achieve this is as follows. A classical solution for directly creating a slow-roll inflating universe was analyzed in Ref.~\cite{Maldacena:2024uhs}, where it was found (see also~\cite{Hertog:2023vot}) that the complex metric involved violates the Kontsevich--Segal--Witten (KSW) criterion~\cite{Louko:1995jw,Kontsevich:2021dmb,Witten:2021nzp} if
\begin{equation}
  \epsilon \leq \frac{1}{2{\cal N}},
\label{eq:KSW}
\end{equation}
where $\epsilon$ is a slow-roll parameter, and ${\cal N}$ denotes the number of $e$-folds of inflation occurring from the point of creation.%
\footnote{
 In certain regimes, such as large-field inflation, a more precise expression is required; see~\cite{Janssen:2024vjn}. Our conclusion below persists if the relevant portion of the potential does not admit such a regime, or if the cosmic histories allowed by the KSW criterion involve only short inflation and, thus, are incompatible with the emergence of structure, as discussed below.
}
This allows us to address the issue by conjecturing that
\begin{itemize}
\item[(1$^\prime$)]
Complex metrics violating the KSW criterion do not contribute to the gravitational path integral, at least at the level of saddle-point approximation.
\end{itemize}
This resolves the curvature problem for two reasons. First, it is likely that a period of very short inflation, ${\cal N} < {\cal N}_*$, does not allow for the emergence of structures capable of supporting observers~\cite{Garriga:1998px,Freivogel:2005vv,Guth:2012ww}. Second, for ${\cal N} > {\cal N}_*$, the potential can have a sufficient slope to violate the criterion of Eq.~(\ref{eq:KSW}) while remaining in the slow-roll regime.

This allows us to rule out the possibility of creating a slow-roll inflating universe with a strong preference for large spatial curvature, while still obtaining sufficient inflation if appropriate initial conditions are provided by another process. The fact that slow-roll inflation in our past is driven by a potential with a reasonable slope fits comfortably with the picture in which the flatness of the potential is required by environmental selection, rather than determined by a theoretical mechanism such as symmetry~\cite{Guth:2013sya}.

\subsubsection*{Problem of direct creation of universes without cosmological evolution}

\begin{figure}[t]
\centering
\vspace{0.3cm}
  \includegraphics[height=0.39\textwidth]{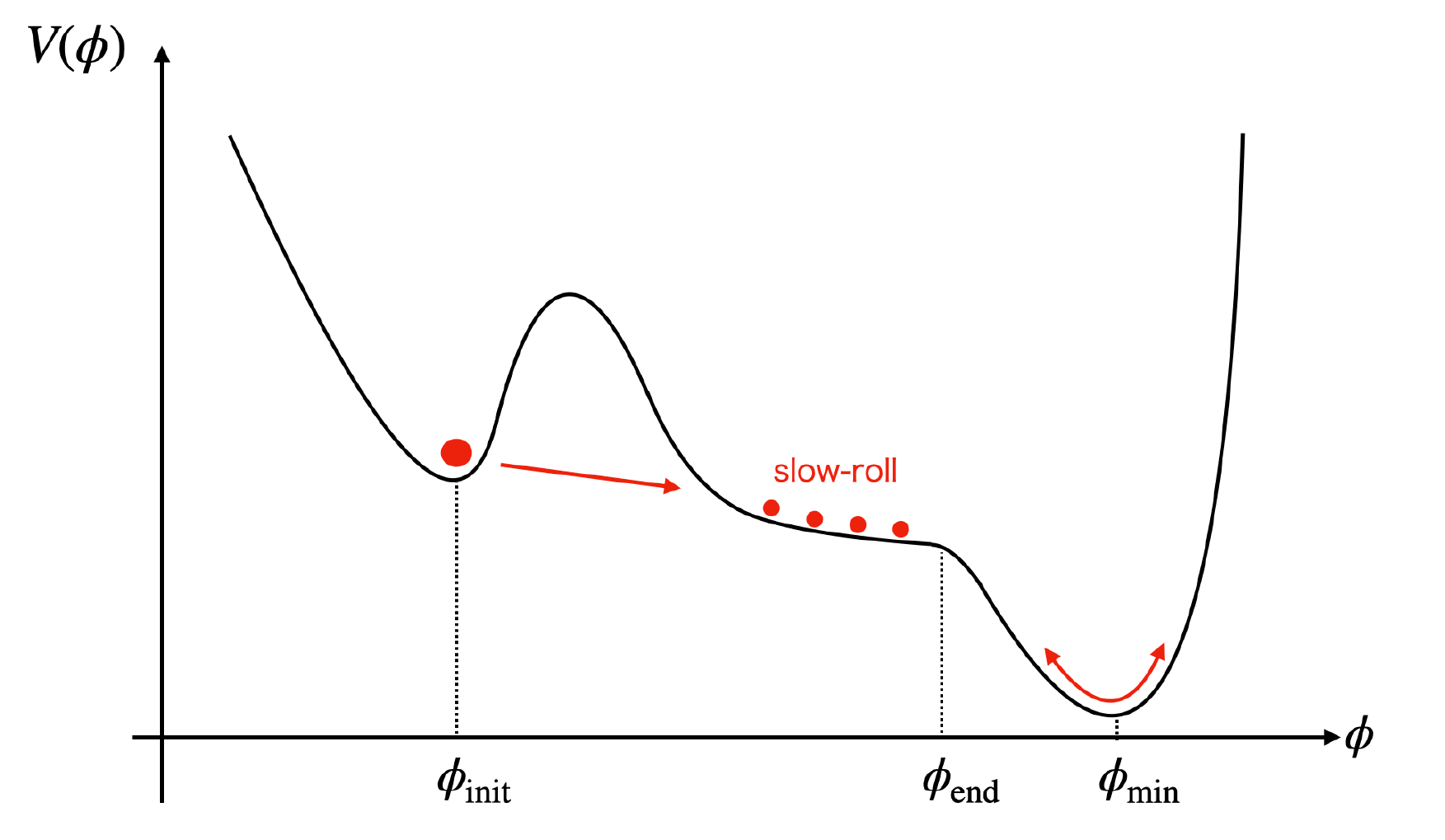}
\vspace{0cm}
\caption{
 Schematic potential for the scalar field $\phi$. The universe is assumed to be created at a metastable local minimum $\phi_{\rm init}$, located above the endpoint of slow-roll inflation $\phi_{\rm end}$. The field subsequently tunnels out of $\phi_{\rm init}$ through bubble nucleation, leading to slow-roll inflation inside the bubble. For a viable cosmology, creation at other points in the potential must be suppressed by a combination of dynamical effects and environmental selection.
}
\label{fig:potential}
\end{figure}

Once the possibility of directly creating a slow-roll inflating universe is eliminated, a successful cosmology would be obtained if the universe is created at a metastable local minimum of the potential, $\phi_{\rm init}$, located sufficiently above the endpoint of slow-roll inflation, $\phi_{\rm end}$; see Fig.~\ref{fig:potential}.
With this initial condition, the standard cosmology with slow-roll inflation follows after the universe tunnels out from $\phi_{\rm init}$ through bubble nucleation~\cite{Coleman:1980aw}. For this scenario to be viable, creation at other points in the potential must be suppressed relative to $\phi_{\rm init}$, due to some combination of dynamical effects and environmental (anthropic) selection, the latter requiring that the resulting universe admit reasonable {\it physical} observers.

It is not difficult to imagine that the creation at $\phi_{\rm init}$ dominates over creations at other minima under the anthropic condition, except for the potential minimum corresponding to our current universe, $\phi_{\rm min}$. It is well known that the Hartle--Hawking factor, i.e.\ the contribution associated with a yarmulke singularity, leads to an overwhelmingly large probability $P(\phi_{\rm min})$ of creating universes directly at $\phi_{\rm min}$, compared with that at $\phi_{\rm init}$
\begin{equation}
  \frac{P(\phi_{\rm min})}{P(\phi_{\rm init})} \sim e^{10^{120}}.
\label{eq:direct-univ}
\end{equation}
While the universe created in this way is empty, the standard argument says that thermal fluctuations associated with the de~Sitter nature of the minimum lead to fluke observers (Boltzmann brains) who find themselves in disordered universes. Because of the huge enhancement factor in Eq.~(\ref{eq:direct-univ}), such observers completely dominate ordinary observers, thereby jeopardizing the viability of the theory~\cite{Dyson:2002pf,Albrecht:2002uz,Bousso:2006xc,Page:2006hr}.

It is possible, however, that the thermal nature of a de~Sitter vacuum does not imply the existence of Boltzmann brains~\cite{Boddy:2014eba,Nomura:2015zda}, much like the Unruh radiation in flat space experienced by an accelerating observer~\cite{Unruh:1976db} does not imply that the Minkowski vacuum can ``think.'' In this picture, the thermal nature of a vacuum as seen in a particular reference frame does not mean that ``physical'' particles exist there, but rather means only that other systems interacting with it respond {\it as if} they are immersed in a thermal bath~\cite{Nomura:2018kia,Nomura:2019qps,Langhoff:2021uct,Murdia:2022giv}. For example, the thermal atmosphere of a black hole by itself does not possess independent ``physical'' dynamics, but instead manifests itself through the response of other systems~\cite{Nomura:2018kia,Nomura:2019qps}, for instance when it is probed by an asymptotic flat region at the edge of the zone~\cite{Hawking:1974sw} or by an object located within the zone~\cite{Unruh:1982ic,Brown:2012un}. Another example is density fluctuations induced by cosmic inflation, which become physical when the system is transformed to a non-inflating (big-bang) universe~\cite{Mukhanov:1981xt,Hawking:1982cz}, or when a material detector is placed in the inflating universe~\cite{Gibbons:1977mu}.

Summarizing, another requirement to obtain a realistic cosmology in the present framework is
\begin{itemize}
\item[(2)]
Boltzmann brains do not arise in an empty de~Sitter vacuum.
\end{itemize}
As discussed in Ref.~\cite{Nomura:2015zda}, this is a statement about the dynamics of the underlying fundamental theory of quantum gravity.

\subsubsection*{Problem of normalizability}

The current framework produces relative probabilities between different observational events. While this may be sufficient as a physical theory, it would be more comfortable if we could discuss a notion of ``absolute probabilities'' in the universe. For this to be the case, the quantum gravity state represented in our unconstrained Hilbert space must be normalizable.

With our assumptions above, it is possible to contemplate that this is indeed the case. First, the absence of Boltzmann brains---assumption~(2)---implies that an indefinitely long-lived de~Sitter vacuum (possibly up to the Poincar\'{e} recurrence time~\cite{Dyson:2002pf}) does not lead to Boltzmann-brain domination. With assumption~(1), the history relevant for physical observers can thus be as follows. The universe is created dominantly at some minimum of the potential higher than the endpoint of slow-roll inflation, with exponentially suppressed probability of being created at even higher minima. After creation, the universe undergoes bubble nucleation and, after a brief period of curvature domination, enters a phase of slow-roll inflation. The standard big-bang cosmology then follows, and the universe eventually ends up in the de~Sitter minimum we currently live in.%
\footnote{
 For consistency, this scenario also requires that any upward tunneling~\cite{Hawking:1981fz,Weinberg:2006pc} from our current vacuum, if it occurs, does not populate a slow-roll inflating regime of the potential. Otherwise, such processes could reintroduce the too-large-curvature problem discussed above. This requirement would be satisfied if upward tunneling connects only stationary points of the potential, reflecting the structure of admissible Euclidean saddle points in the gravitational path integral.
}

This scenario is consistent with the picture~\cite{Guth:2013sya} that a slow-roll inflating phase is required to obtain physical observers. After tunneling, the universe is initially dominated by curvature, whose contribution scales as $1/a^2$, where $a$ is the scale factor. Since matter, radiation, and kinetic contributions redshift more rapidly (as $1/a^3$, $1/a^4$, and $1/a^6$, respectively), the universe must pass through a region of a reasonably flat potential, allowing the inflaton energy (scaling roughly as $1/a^0$) to dominate and subsequently convert into radiation or matter energy density, making the existence of observers possible.

The probabilities in such a scenario will be normalizable because the dimension of the relevant Hilbert space is expected to be finite in a universe with a positive vacuum energy~\cite{Banks:2000fe,Witten:2001kn}. We emphasize that this scenario most likely lies in a regime beyond semiclassical physics. From a semiclassical perspective, if there exists any vacuum with zero or negative energy density, then our current vacuum would be able to decay into such vacua, rendering the probabilities nonnormalizable due to the existence of an infinite number of possible late-time configurations. The state must therefore be {\it selected} by the requirement of normalizability, much like the bound states of the electron in the hydrogen atom are selected~\cite{Nomura:2012zb}; from the semiclassical point of view, these atomic states appear as infinitely fine-tuned configurations among an infinite number of possible (nonnormalizable) asymptotic electron states.

Just as the bound states of the hydrogen atom are difficult to analyze using a basis of asymptotic scattering states, the cosmological state may also be difficult to analyze using formulations of quantum gravity based on asymptotic states, such as the $S$-matrix formulation of string theory~\cite{Polchinski:1998rq,Polchinski:1998rr} or the anti--de~Sitter/conformal field theory correspondence~\cite{Maldacena:1997re}. While these formulations are powerful in uncovering physics at sub-Planckian length scales, their direct applicability to cosmology is not obvious. Extracting cosmological physics might still require an effective description of quantum gravity of the type considered in this paper.

\section*{Acknowledgments}

We are grateful to the Yukawa Institute for Theoretical Physics (YITP), Kyoto University, for its hospitality during the workshop YITP-T-25-01, where part of this work was done. This work was supported in part by MEXT KAKENHI Grant Number JP25K00997. The work of Y.N. was also supported by the U.S. Department of Energy, Office of Science, Office of High Energy Physics under QuantISED Award DE-SC0019380 and Contract No.\ DE-AC02-05CH11231, and by the Japan Science and Technology Agency (JST) as part of Adopting Sustainable Partnerships for Innovative Research Ecosystem (ASPIRE), Grant No.\ JPMJAP2318.

\end{document}